\documentclass[preprint2]{aastex}

\slugcomment{To be submitted to the ApJ}

\shorttitle{Lupus V and VI synthesis paper}
\shortauthors{Spezzi et al.}

\begin{document}

\title{The Spitzer Survey of Interstellar Clouds in the Gould Belt. IV. \\ Lupus V and VI Observed with IRAC and MIPS}

\author{Loredana Spezzi\altaffilmark{1},
Pierre Vernazza\altaffilmark{1},
Bruno Mer\'{\i}n\altaffilmark{2},
Lori E. Allen\altaffilmark{3},
Neal J. Evans II\altaffilmark{4},
Jes K. J{\o}rgensen\altaffilmark{5},
Tyler L.  Bourke\altaffilmark{6},
Lucas A. Cieza\altaffilmark{7},
Michael M. Dunham\altaffilmark{8},
Paul M. Harvey\altaffilmark{4},
Tracy L. Huard\altaffilmark{9},
Dawn Peterson\altaffilmark{6},
Nick F. H. Tothill\altaffilmark{10},
\& the Gould's Belt Team
}

\altaffiltext{1}{Research and ScientiÞc Support Dept., ESTEC (ESA), Keplerlaan, 1, PO Box 299, 2200 AG Noordwijk, The Netherlands}
\altaffiltext{2}{Herschel Science Centre, ESAC (ESA), P.O. Box 78, 28691 Villanueva de la Ca\~nada (Madrid), Spain}
\altaffiltext{3}{Department of Astronomy, University of Arizona, 933 N Cherry Ave., Tucson AZ 85721-0065, USA}
\altaffiltext{4}{Department of Astronomy, University of Texas at Austin, 1 University Station C1400 Austin, TX 78712-0259, USA}
\altaffiltext{5}{Centre for Star and Planet Formation, Natural History Museum of Denmark, University of Copenhagen, {\O}ster Voldgade 5-7, DK-1350 Copenhagen, Denmark} 
\altaffiltext{6}{Harvard-Smithsonian Astrophysical Observatory, 60 Garden Street, Cambridge, MA 02138, USA}
\altaffiltext{7}{Institute for Astronomy, University of Hawaii at Manoa, Honolulu, HI 96822, USA}
\altaffiltext{8}{Department of Astronomy, Yale University, P.O. Box 208101, New Haven, CT 06520, USA}
\altaffiltext{9}{Astronomy Department, University of Maryland, College Park, MD 20742, USA}
\altaffiltext{10}{School of Physics, University of Exeter, Exeter, EX4 4QL, UK}

\begin{abstract}

We present Gould's Belt (GB) Spitzer IRAC and MIPS observations of
the Lupus V and VI clouds and discuss them in combination with
near-infrared (2MASS) data. Our observations complement those
obtained for other Lupus clouds within the frame of the
Spitzer ``Core to Disk'' (c2d) Legacy Survey. We found 43 Young
Stellar Object (YSO) candidates in Lupus V and 45 in Lupus VI, including 2 transition disks, 
using the standard c2d/GB selection method. None of these sources was 
classified as a pre-main sequence star from previous optical, near-IR and X-ray surveys. 
A large majority of these YSO candidates appear to be
surrounded by thin disks (Class III; $\sim$79 \% in Lupus V and $\sim$87\% in Lupus VI). These Class III abundances differ
significantly from those observed for the other Lupus clouds and c2d/GB surveyed star-forming regions, 
where objects with optically thick disks (Class II) dominate the young population. 
We investigate various scenarios that can explain this discrepancy. In particular, we
show that disk photo-evaporation due to nearby OB stars is not responsible for the high fraction of
Class III objects. The gas surface densities measured for Lupus V and VI lies below the star-formation threshold (A$_V \approx$8.6~mag), 
while this is not the case for other Lupus clouds. Thus, few Myrs older age for the YSOs in Lupus V and VI with 
respect to other Lupus clouds is the most likely explanation of the high fraction of Class III objects in these clouds, 
while a higher characteristic stellar mass might be a contributing factor.  
Better constraints on the age and binary fraction of the Lupus clouds might solve the puzzle but require further observations.

\end{abstract}


\section{Introduction \label{intro}}

Circumstellar disks of gas and dust, which surround the vast
majority of young stars, are the reservoirs of material out of
which planets may form \citep{Lada1992, Kenyon1995, Currie2009}.
Their presence can be inferred via spectroscopic and/or
photometric observations at infrared wavelengths as emission in
excess of the stellar photosphere. The incidence of primordial
protoplanetary accretion disks (with inner radii within 0.1 AU of
the central star and a strong, optically thick emission)
diminishes with age, being very common at ages of less than 1 Myr
and very rare at more than 10 Myr \citep{Mamajek2004}. Stars older
than 10 Myr usually show no excess emission, meaning no disk, or have an optically thin mid-infrared emission
(tracing material between 0.3 and 3 AU around typical low-mass stars) and show little evidence
for substantial reservoirs of circumstellar gas, indicating that
they are surrounded by debris disks \citep[e.g.,][]{Rieke2005,Dahm2005,Currie2008a,Currie2008b,Currie2009,Hillenbrand2008}. 
Because dust from debris disks can be
removed by stellar radiation on very short timescales (less than
0.1 Myr), the presence of dust requires an active replenishment source from
collisions between larger objects \citep[e.g.,][]{Burns1979,Plavchan2005}. 
Thus, debris disks around young stars are an indication of
active planet formation and their ages also place an upper limit
on the formation timescale of gas giant planets. Determining the
timescale for the disappearance of primordial disks, and the
subsequent dominance of debris disks, is then of primary importance \citep{Currie2009}.

The latter goal has been one of the key objectives of the Spitzer
Legacy Project ``From Molecular Cores to Planet-forming disks'' 
\citep[c2d; see][]{Evans2003}. The c2d observations included partial
spatial coverage of the following clouds: i) Lupus I, III and IV
\citep{Merin2008}, ii) Cha II \citep{Alcala2008}, iii) Perseus \citep{Evans2009a}, iv) Serpens \citep{Harvey2007b}, and
v) Ophiuchus \citep{Padgett2008}. 
The results of the c2d survey were combined by \citet{Evans2009a} in order to provide
an updated statistical analysis of the global properties of star formation in all these clouds (star formation rates and
efficiencies, numbers and lifetimes of YSOs in each infrared class, clustering properties, etc.).

The Gould's Belt (GB) project is a continuation of the c2d project to
complete the Spitzer mapping of nearby star-forming regions in the Gould's Belt
(Allen et al., in preparation). In this paper, we present results of the IRAC and MIPS 
observations of the Lupus V and VI clouds. These observations complement
those of \citet{Merin2008} for Lupus I, III, and IV, and allow us
to provide a quite complete overview of the evolutionary status of
the Lupus star-forming complex. The Lupus cloud complex (RA $\sim$ 16$^h$20$^m$ -- 15$^h$30$^m$ and
DEC $\sim$ -43$^{\rm o}$00$'$ -- -33$^{\rm o}$00$'$ ) lies between 150 and 200 pc from the Sun \citep{Comeron2008_rev}, 
at galactic longitudes 334$^{\rm o} < l < 352^{\rm o}$ and latitudes $+5^{\rm o} < b < +25^{\rm o}$ \citep{Krautter1992}. 
Lupus is one of the main low-mass star forming
complexes, with mid M-type pre-main sequence (PMS) stars dominating its stellar population \citep{Hughes1994}. 
The complex is located in the Scorpius-Centaurus
OB association, whose massive stars are likely to have played a
significant role in the evolution and perhaps the origin of the
complex \citep{Comeron2008_rev}. With an age of $\sim$1.5-4 Myr
\citep{Hughes1994, Comeron2003}, Lupus lies between the epoch when
most stars have optically-thick primordial disks ($\sim$1 Myr)
and the epoch where stars are mostly surrounded by debris disks
($\sim$10 Myr). The age estimate strongly relies on the
adopted distance to each Lupus cloud. Because the Lupus complex is spread over about 50~pc along the line of sight, the derived stellar age should 
be considered with caution and apparent age spreads up to 10~Myr or more might be due the uncertain 
location of each cloud between 150 and 200 pc  \citep{Comeron2008_rev}.  
For a detailed review of the Lupus complex we defer the reader to 
\citet{Comeron2008_rev},  \citet{Cambresy1999}, and \citet{Teixeira2005}. 

We organize our paper as follows. We first describe our Spitzer observations and
data reduction followed by a description of the process by which
we identify YSO candidates and eliminate field contaminants. 
We then characterize the global properties of the YSO candidate samples 
and compare our results with those of earlier studies for
the Lupus complex \citep{Merin2008} and of other star forming regions observed by Spitzer \citep{Evans2009a}. 
Finally, we discuss possible evolution scenarios consistent with the observed disk properties of the YSO candidates in Lupus V and VI. 
In Appendix~A we present a detailed description of the several criteria used to corroborate the disk properties of the YSOs candidates.

\section{Spitzer IRAC and MIPS data \label{IRAC_BDP}}

\begin{figure*}
\includegraphics[angle=0,scale=0.9]{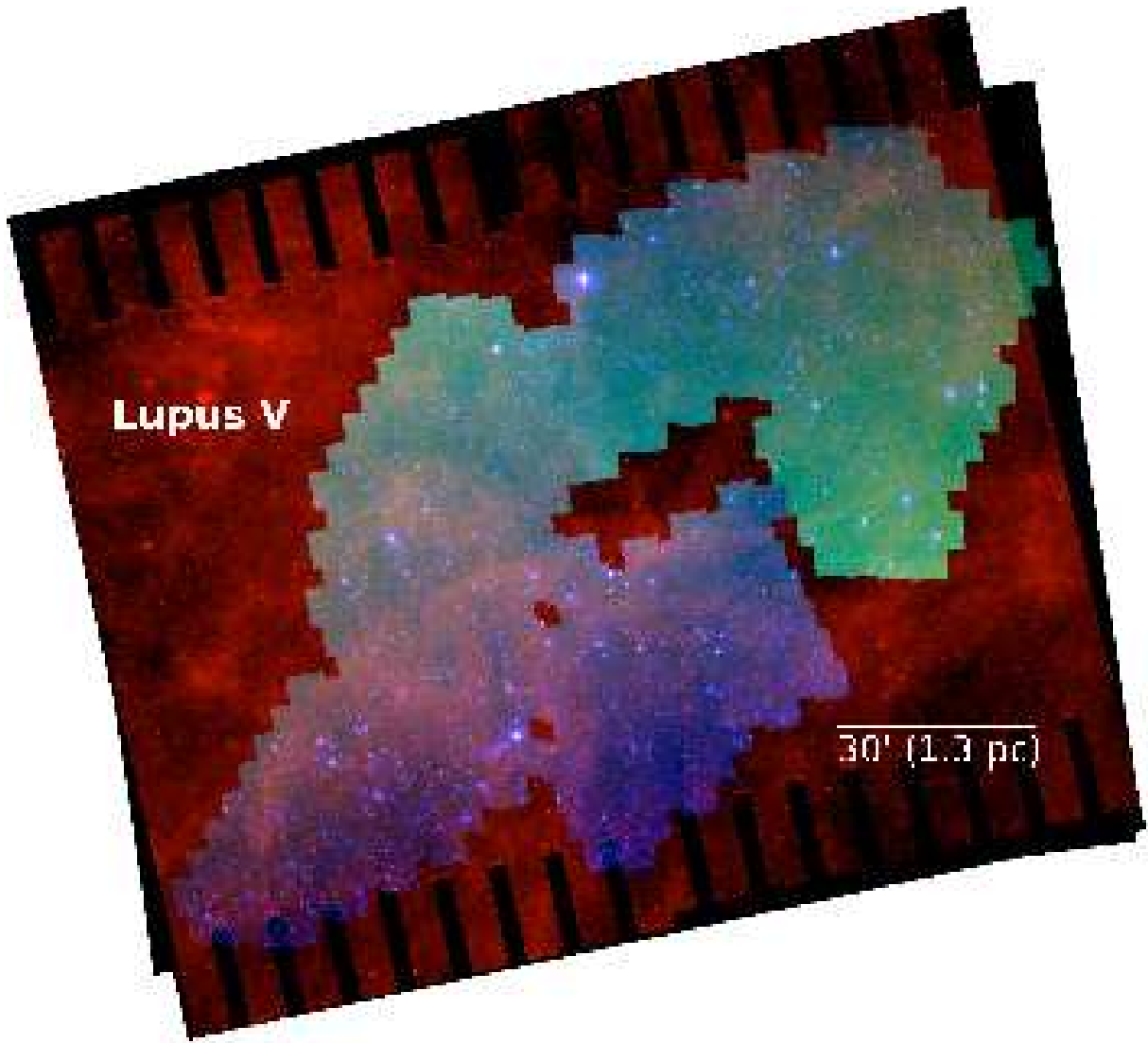}
\caption{Color-composite image of the area in Lupus~V mapped by the GB survey. North is up and East to the left. 
The color code is blue for the IRAC~2 band at 4.5$\mu$m, green for the IRAC~4 band at 8$\mu$m, and red for the MIPS~1 band at 24$\mu$m;  
note that the IRAC~2 and IRAC~4 mosaics cover exactly the same sky area.
The figure shows the remnant cloud structure emitting at long wavelengths. \label{LupV_rgb}}
\end{figure*}

\begin{figure*}
\includegraphics[angle=0,scale=0.9]{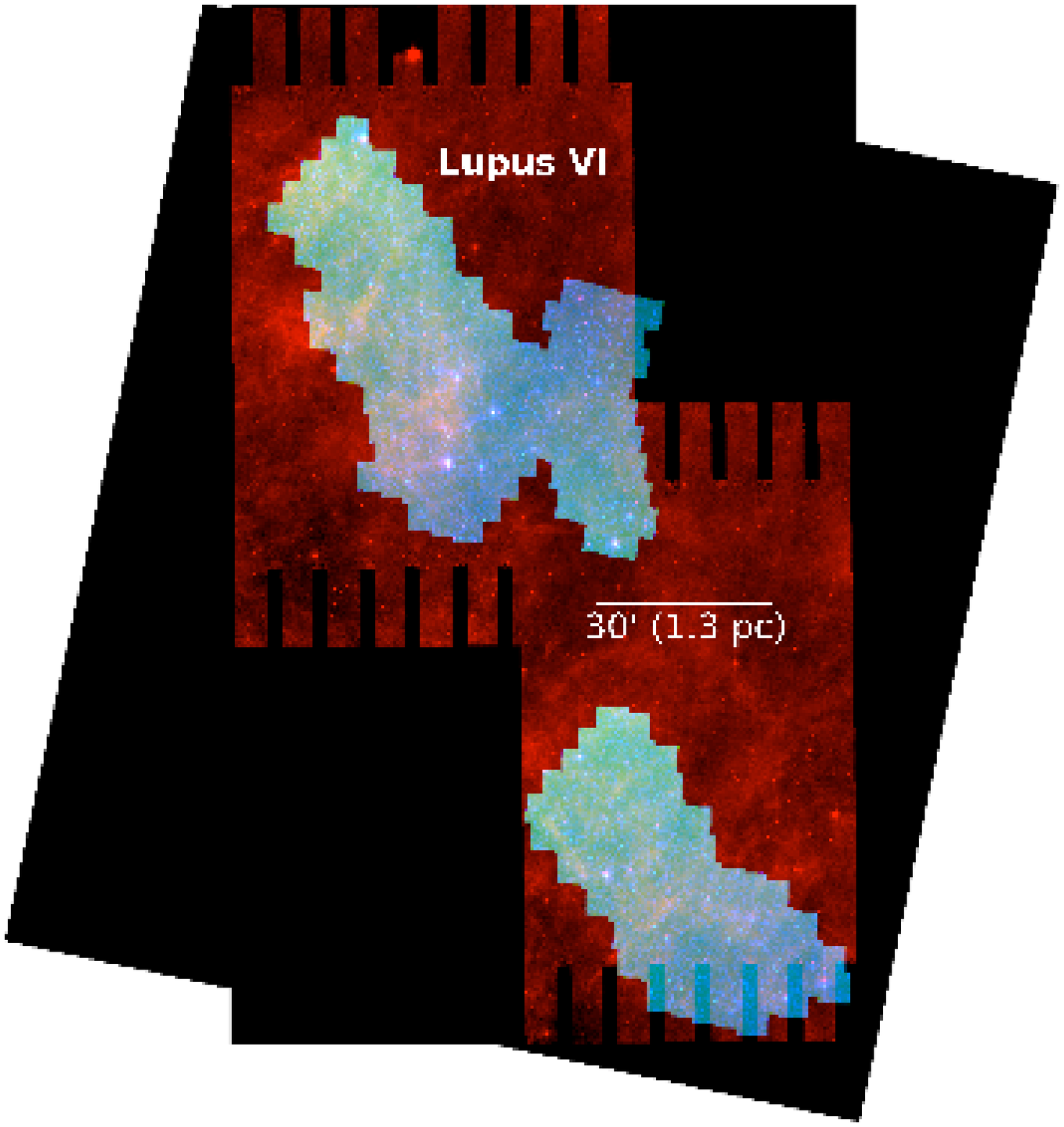}
\caption{Same as Figure~\ref{LupV_rgb} for Lupus~VI. \label{LupVI_rgb}}
\end{figure*}

The Lupus V and Lupus VI dark clouds were observed using the Spitzer Space Telescope IRAC
and MIPS cameras at 3.6, 4.5, 5.8, 8.0, 24 and 70 $\mu$m on 1-8
April and 9 September 2007 as part of the Spitzer GB survey (PI: Lori E. Allen). 
A 3.82~deg$^2$ and 2.88~deg$^2$ areas were mapped in Lupus V and Lupus VI, respectively, covering entirely the cloud area
where the visual extinction is greater than A$_V \approx$2 (Figure~\ref{LupV_rgb} and \ref{LupVI_rgb}) 
in the extinction map reported by \citet{Cambresy1999}. 
The observations in Lupus V and VI were performed to match the detection 
and sensitivity limits of the Spitzer c2d survey in Lupus~I, III and VI. 

The data reduction procedure was essentially identical to
the one used for all the c2d data. The IRAC and MIPS
images were first processed by the Spitzer Science Center (SSC)
using the standard pipeline to produce the Basic Calibrated Data (BCD). The c2d pipeline for IRAC data has been
described by \citet{Harvey2006} and \citet{Jorgensen2006}, while the
pipeline for MIPS data has been described by \citet{Young2005} and
\citet{Rebull2007}. Using these BCD images, the final source
catalogs were produced following four additional steps:
\begin{enumerate}

\item Extra bad-pixel masking, image correction for
muxbleed, column pull-downs/pull-ups as well as correction for
the ``first frame'' effect in the third IRAC band \citep{Harvey2006};

\item Mosaicking of the individual frames. This was done with the SSC's ``Mopex'' software suite
\citep{Makovoz2005}, which eliminates detector transient effects and transient sources (like asteroids) from the mosaics;

\item Source location and extraction from the stacked mosaics, 
using the c2d developed tool {\sl c2dphot} \citep{Harvey2006}.
This software extracts photometry and calculates the uncertainties
using digitized point spread functions. It also gives a 
morphological classification of the extracted sources (point-like or extended);

\item Cross-identification of the sources among the 4 IRAC bands
and merging of the final IRAC catalog with the MIPS 24 and 70~$\mu$m catalogs \citep{Chapman2007} and with the near-IR 2$\mu$m
All Sky Survey (2MASS) catalog \citep{Cutri2003}. For the band
merging, we used a 2.0$''$ match radius among IRAC bands and 2MASS
and 4.0$''$ and 8.0$''$ for matching with MIPS bands 24 and
70$\mu$m, respectively, to account for the larger point spread
functions of the MIPS bands.

\end{enumerate}

For further information on the data reduction procedure we defer the reader 
to the delivery document by \citet{Evans2007}.  Table \ref{detection_stat} summarizes the number of sources
detected with signal-to-noise ratio of at least 5 in each band in the two surveyed areas in Lupus V and VI. Note that most of the detected
objects are background giant or foreground dwarf stars at the
brighter flux levels, while extragalactic objects make a
substantial contribution to the fainter source counts.

\section{Selection of YSO candidates \label{yso_sel}}

The main goal of the GB observations is the selection of YSO candidates based on their IR excess emission with respect to older field stars. 
However, the IR colors of many galaxies are very similar to those of YSOs. 
Thus, a sample of YSO candidates must be corrected for the contamination of background extragalactic objects. 
To distinguish YSOs from galaxies, we used a combination of 2MASS, IRAC and MIPS color-color
(CC) and color-magnitude (CM) diagrams. This method has been
developed within the frame of the Spitzer c2d Legacy Survey and
has proven to be a success. Indeed, most of YSO candidates selected in Cha~II, Lupus~III and Serpens have been subsequently observed via spectroscopy 
and the YSO nature has been confirmed for $\sim$70\% of them \citep[see, e.g.,][]{Spezzi2008,Merin2008,Oliveira2009,Cieza2010}. 

A detailed review of the selection method can be found in
\citet{Harvey2007a,Harvey2007b}. Briefly, the selection method consists in the definition of an
empirical probability function which depends on the relative
position of a given source in several CC and CM diagrams, where diffuse boundaries have been
determined to obtain an optimal separation between young stars and
galaxies. 

\begin{figure*}
\includegraphics[angle=0,scale=0.9]{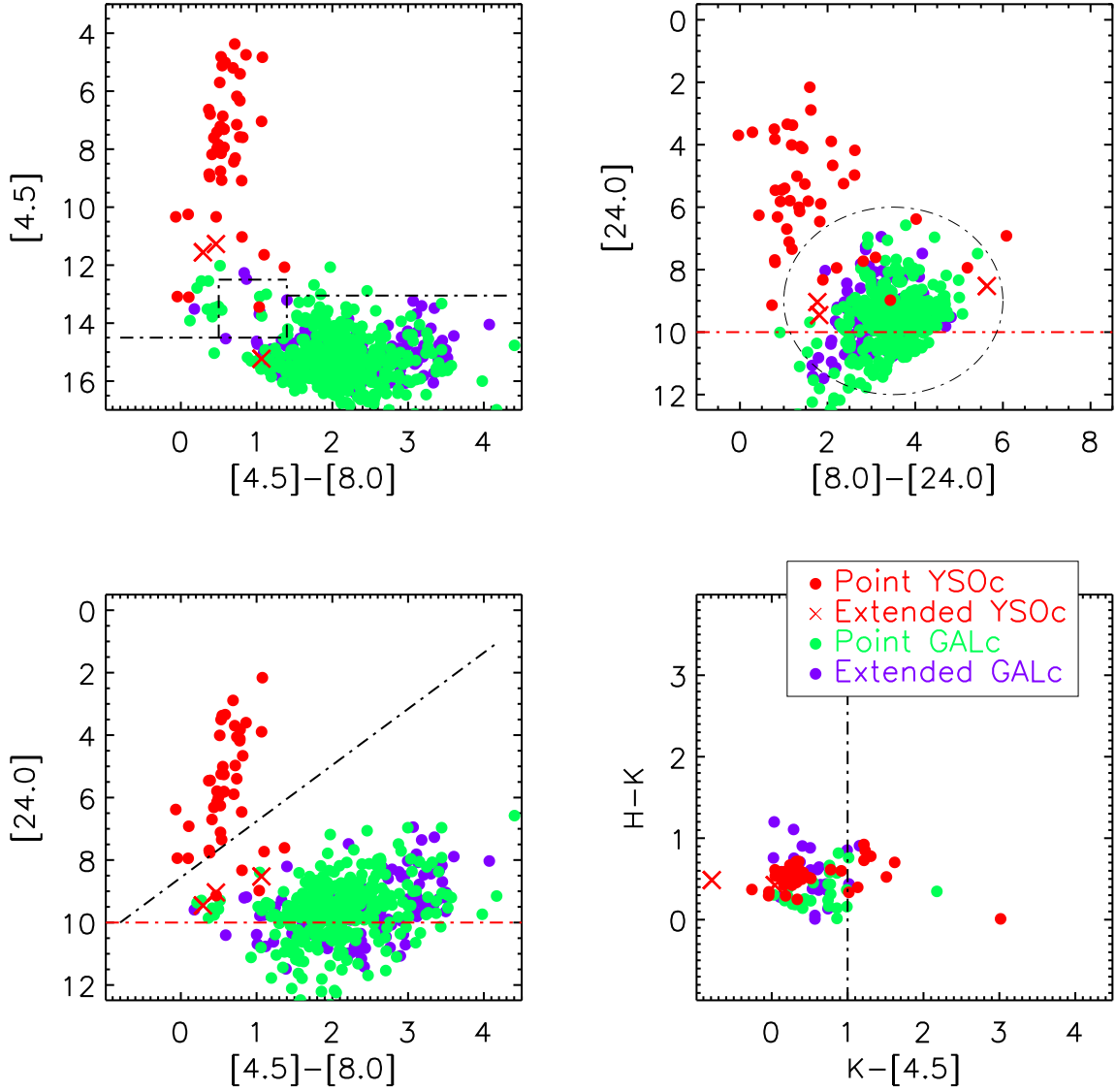}
\caption{CM and CC diagrams for Lupus V. The black
dot-dashed lines show fuzzy limits with exponential cut-offs that defines the YSO candidate (YSOc) selection 
criterion in the each diagram, excluding contamination from galaxy candidates (GALc). 
Field stars presenting normal photospheric colors are not plotted. 
The red dot-dashed lines show hard limits, fainter than which objects are
excluded from the YSO category. Symbols are as in the legend. \label{LupV_sel}}
\end{figure*}

\begin{figure*}
\includegraphics[angle=0,scale=0.9]{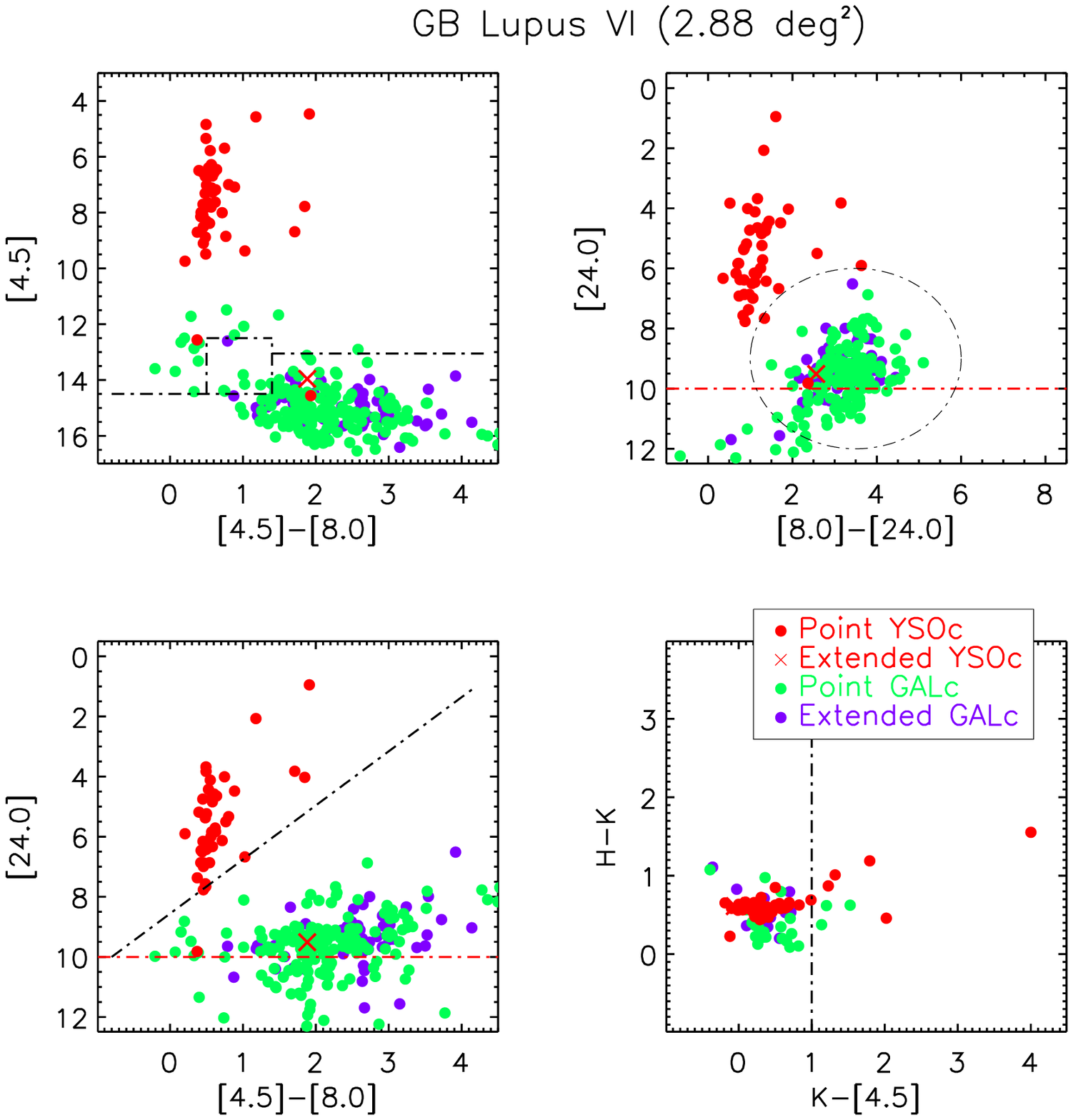}
\caption{Same as Figure~\ref{LupV_sel} for Lupus VI. \label{LupVI_sel}}
\end{figure*}

Figures~\ref{LupV_sel} and \ref{LupVI_sel} show the Spitzer CC and CM diagrams used to select
the YSO candidates in Lupus V and VI, respectively. Note that the method
requires detection in all IRAC bands and in MIPS1 with a S/N
higher than 3 to classify an object as a YSO candidate or a background
galaxy.  Moreover, older field objects with no IR excess emission are rejected \emph{a priori} 
because their IR colors are comparable with normal photospheric colors \citep[e.g., $K-4.5<$-0.1;][]{Harvey2007a}. 
We find 43 YSO candidates in Lupus V and 45 in Lupus VI, shown in Figures~\ref{LupV_sel} and
\ref{LupVI_sel} as red dots and crosses for point-like and
extended sources, respectively. All the YSO candidates have been visually inspected in the IRAC and MIPS images; they appear point-like in all our images and their 
photometry is not contaminated by crowding, nearby saturated stars or any other artifact that might affect our selection criterion.  
A remaining caveat is that our YSO candidates might be members of binary/multiple systems too close to be resolved with IRAC/MIPS and, hence, 
affecting the measured photometry. The multiplicity fraction for low mass stars (0.5-1~M$_{\odot}$) is estimated to be between 20\% and 40\%, 
depending on the actual mass of the primary star and the separation range 
\citep{Duquennoy1991,Mason1998,Basri2006,Lada2006a}. 
Higher resolution photometry or spectroscopy would be needed to assess the actual multiplicity fraction in Lupus V and VI. 
The effect of multiplicity on the disk properties of YSOs are discussed in  Sect.~\ref{binary}.

Interestingly, none of these 88 sources in Lupus V and VI was classified as a pre-main sequence star from previous ground-based optical/near-IR
observations and none of them was detected by the ROSAT X-ray survey \citep{Krautter1997} or by the IRAS satellite.
This ``non-detection/classification" is most likely due to photometric incompleteness and lack of dedicated observations for these two clouds. 
Indeed, no deep optical/near-IR surveys, comparable to those conducted in other Lupus clouds \citep{Comeron2009}, are available so far for Lupus V and VI. 
Prior to our Spitzer GB observations, only 5 far-IR sources were identified in these two clouds by the IRAS satellite \citep[see Table~5 by][]{Comeron2008_rev}. 
As for X-ray surveys, the ROSAT all sky survey detected about 200 weak-line T~Tauri stars in the Lupus complex mainly concentrated around  Lupus~III \citep{Krautter1997}. 
Only an handful of them ($\sim$10) are associated with Lupus~V and VI, but none of these X-ray sources overlaps with our YSO candidates. 
No dedicated X-ray observations with XMM or Chandra have been conducted in these two clouds.

In Table~\ref{flux_LupV}-\ref{flux_LupVI} we list IRAC/MIPS fluxes of the YSO candidates selected in Lupus V and VI, 
while in Table~\ref{tab_mag_LupV}-\ref{tab_mag_LupVI} we report their 2MASS magnitudes.

\section{Properties of the YSO candidates \label{YSO_prop}}

\subsection{Spectral Energy Distributions and IR Class  \label{SEDs}}

For each of the YSO candidates selected in Lupus V and VI we
constructed the spectral energy distribution (SED), which can be
used to constrain specific physical properties of the YSO population to be
compared with those of other young populations observed with Spitzer. 
For all objects, the GB catalog provides the 2MASS near-IR
magnitudes and the IRAC~3.6-8$\mu$m and MIPS~24-70$\mu$m fluxes.
We complemented these data with optical photometry (BVRI) from the
NOMAD \citep{Zacharias2005} and DENIS \citep{Denis2005} catalogs,
thus obtaining complete SEDs from optical to infrared wavelengths depending on available data. 
Figures \ref{seds_II}-\ref{seds_IIfin} show the SEDs for all YSO candidates in Lupus V and Lupus VI. 
In each plot we also show, only for visual comparison purposes, 
the NextGen model spectrum corresponding to a K7 stellar photosphere \citep{Hauschildt1999} 
normalized to the J-band dereddened flux. To perform this comparison, we roughly estimate the visual extinction 
toward each YSO candidate by fitting the observed photometry between V and J
with the NextGen model spectrum, arbitrarily assuming a K7 spectral type and $\log g=$4, 
e.g. the typical values of T Tauri stars in Taurus \citep[see][]{Hartmann2005}. The dereddened 
flux at each pass-band was obtained adopting the extinction law 
by \citet{Weingartner2001} with $R_{V}$=5.5, also adopted for other Lupus clouds \citep{Merin2008}. 
Each plot in Figures \ref{seds_II}-\ref{seds_IIfin} also shows, for comparison, the
typical SED of an optically thick accreting disk, i.e. the median SED of T Tauri stars in
Taurus \citep{DAlessio1999} normalized to the dereddened J-band flux. 
From these comparisons, it is clear that the
SEDs of our YSO  candidates show considerable variety. An ideal
classification system has to rely on disk models, which are 
degenerate in several parameters \citep{Robitaille2007}, and also
requires a good knowledge of the physical parameters of the central
star (spectral type, reddening, etc), which are not available from
the literature for our YSO candidate sample. Therefore, we do not attempt a
full disk characterization of each SED but rather classify our objects on the basis of several SED slopes, 
following the approach adopted in other c2d/GB papers \citep{Harvey2007a,Merin2008,Alcala2008}.

\begin{figure*}
\includegraphics[angle=0,scale=0.9]{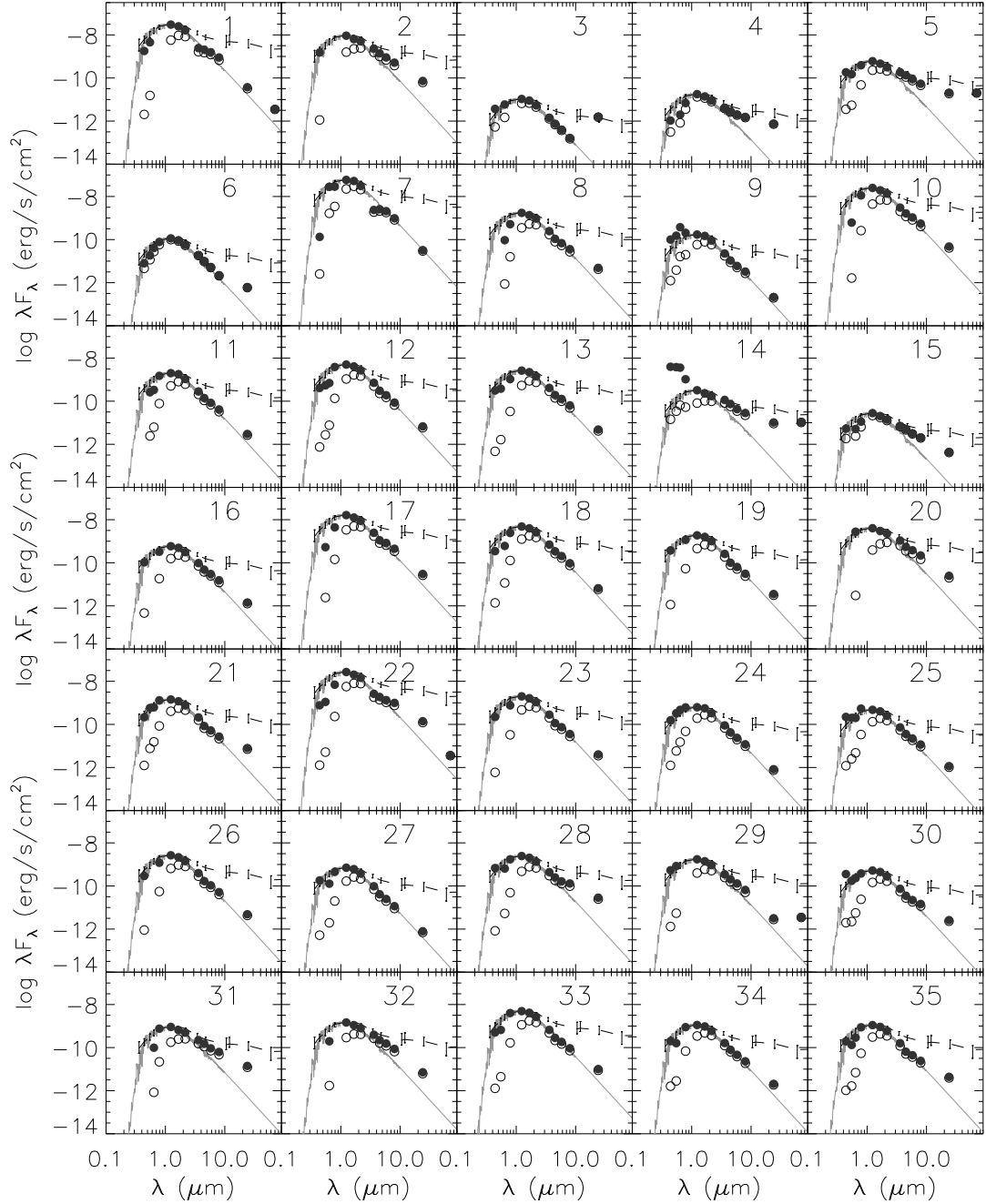}
\caption{Observed (open circles) and dereddened (filled circles)
SEDs of the YSO candidates in the Lupus V cloud. The continuous
line shows the NextGen photospheric emission model of a K7-type star; the dashed line is the average SED of
T~Tauri stars in Taurus. Both these reference models are normalized to the J-band flux. \label{seds_II}}
\end{figure*}

\begin{figure*}
\includegraphics[angle=0,scale=0.9]{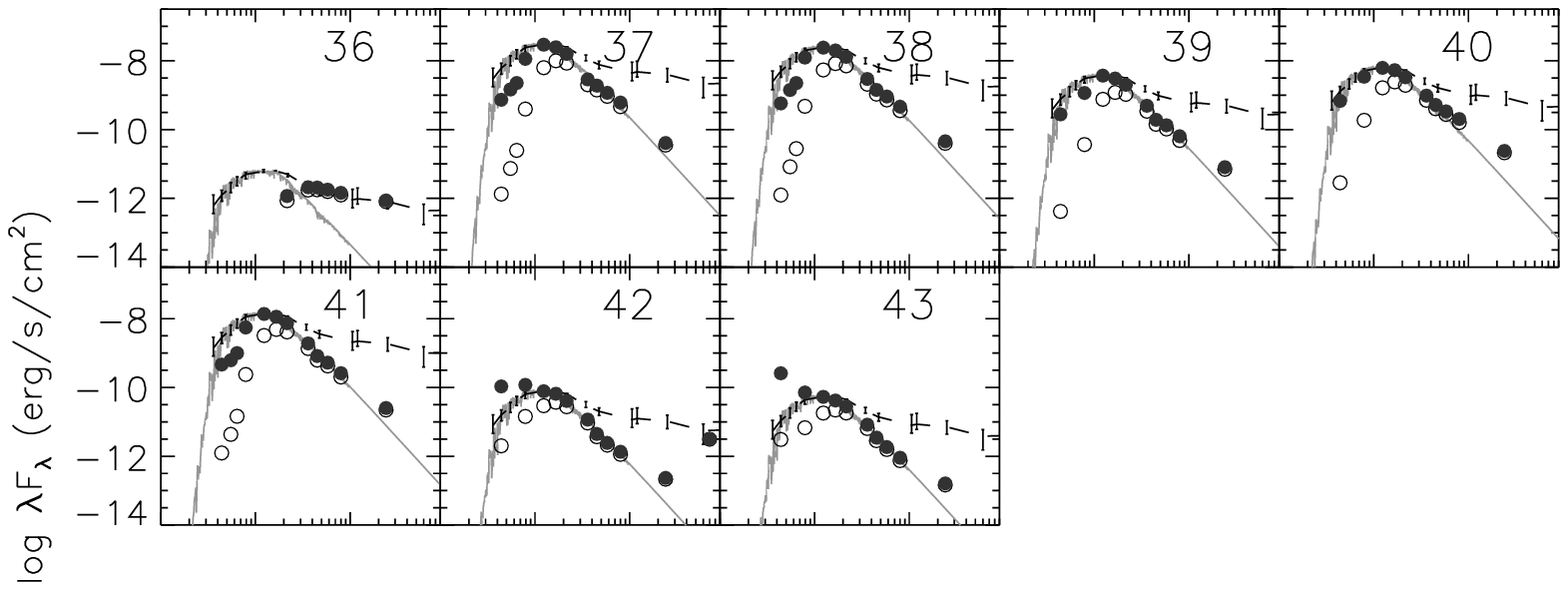}
\caption{Continued.}
\end{figure*}

\begin{figure*}
\includegraphics[angle=0,scale=0.9]{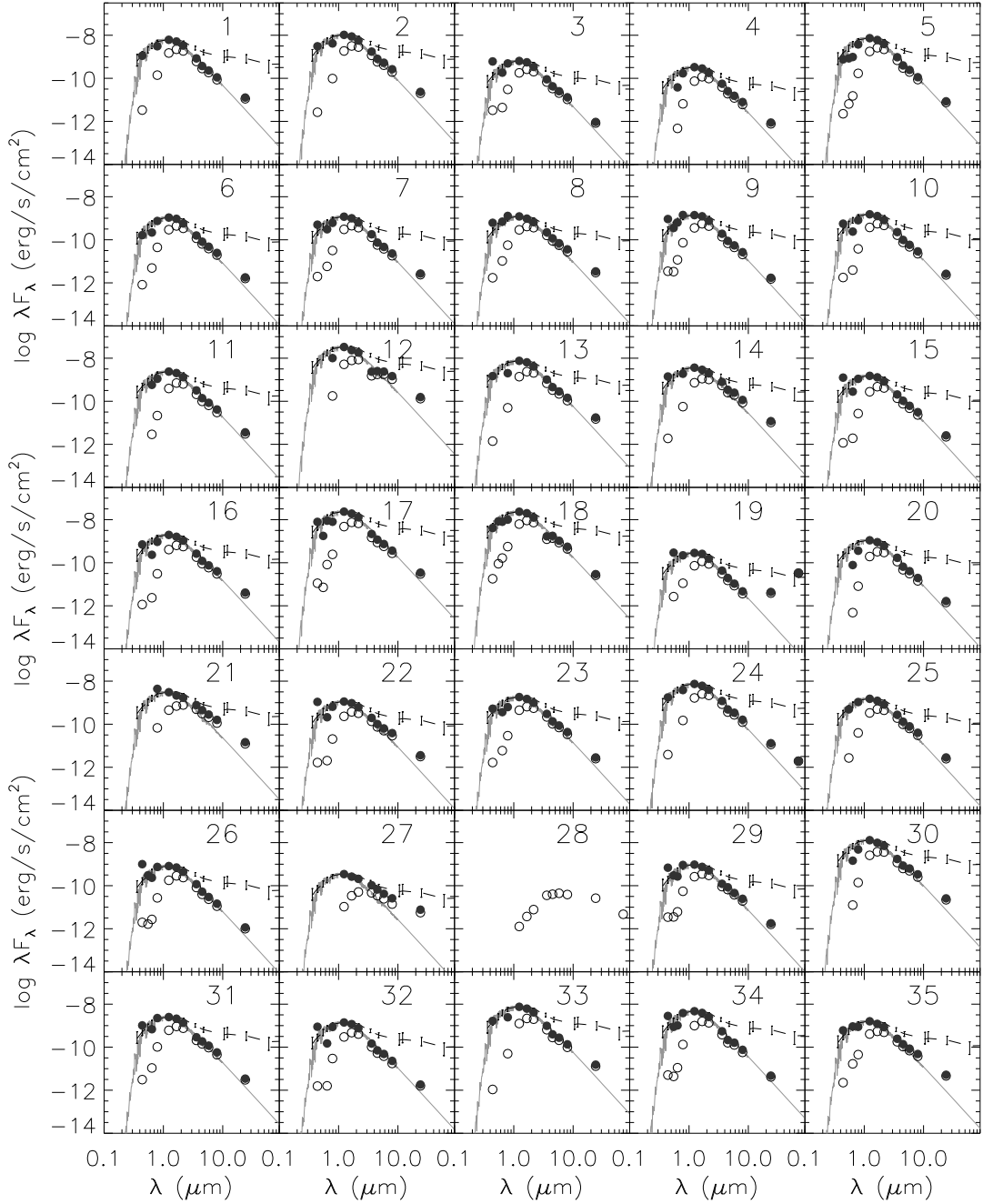}
\caption{Same as Figure~\ref{seds_II} for Lupus VI. The flat-spectrum YSO candidate (ID~28) is not detected in the optical and, hence, 
we do not attempt the comparison with the K7-type stellar model and the  the average SED of T~Tauri stars in Taurus (see Sect.~\ref{flat_obj}. \label{seds_VI}}
\end{figure*}

\begin{figure*}
\includegraphics[angle=0,scale=0.9]{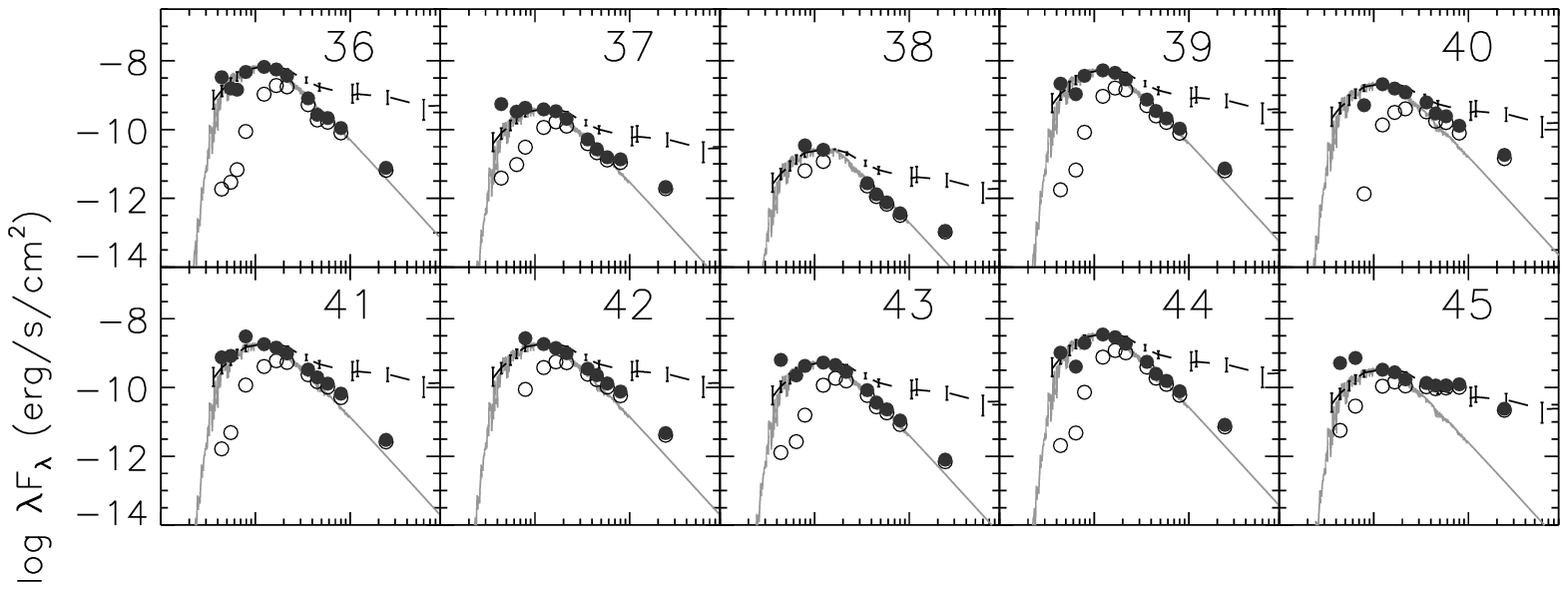}
\caption{Continued. \label{seds_IIfin}}
\end{figure*}

\begin{figure*}
\centering
\includegraphics[angle=0,scale=0.4]{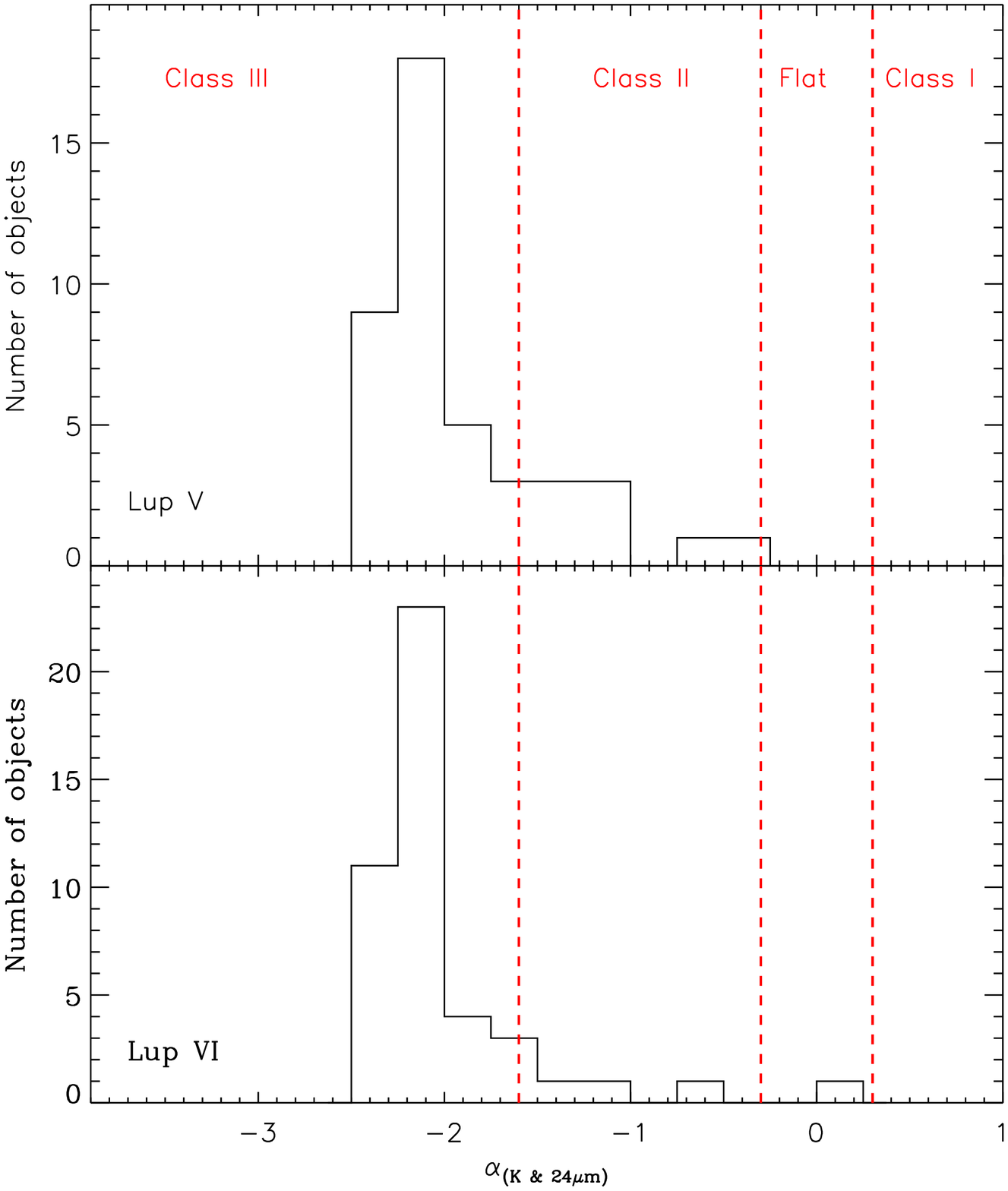}
\caption{$\alpha_{K-24}$-slope distribution of YSO candidates in Lupus V and
Lupus VI (solid line). The vertical dashed lines indicate the
intervals defining the four Lada classes. Both populations are
largely dominated by Class III objects. \label{alpha}}
\end{figure*}

We grouped the YSO candidates identified in Lupus V and VI in four IR
classes (Class I, Flat, Class II and Class III) following the
scheme proposed by \citet{Lada1984} and \citet{Greene1994}. 
Such grouping allows distinguishing between stars with optically thick
and optically thin disks\footnote{In this paper we use the definition ``optically thin disk'' to indicate 
a disk optically thin at most radii to the radiation from the star. This kind of disk is usually associated with a disk-to-stellar 
luminosity ratio of $L_{disk}/L_{star}<$0.01 \citep{Evans2009b}.}; overall, such classification gives a good
overview of the evolutionary state of YSOs within a cloud. This
classification uses the spectral slope ($\alpha$) of the SED at
wavelengths longer than 2$\mu$m; in our case, we computed the
spectral slope of each SED  by performing a least squares fit of all available data in the K-24 $\mu$m range and 
divided our sample in the four classes according to the following
criteria: Class~I $\Rightarrow$ $\alpha \ge$0.3, Flat
$\Rightarrow$ -0.3$\le \alpha <$0.3, Class~II $\Rightarrow$
-1.6$\le \alpha <$-0.3, Class~III $\Rightarrow$ $\alpha <$-1.6. 
The values of $\alpha$-slope for the YSO candidates Lupus V and VI are reported in
Table~\ref{tab_param_V} and \ref{tab_param_VI}. Figure~\ref{alpha} shows the
distribution of spectral slopes for both clouds and
Table~\ref{classes} lists the number of sources in each of the
four classes. From Table~\ref{classes} and Figure~\ref{alpha} 
we can immediately see that both clouds contain a high fraction of
Class III objects (79\% for Lupus V and 87\% for Lupus VI), while
all other clouds observed with Spitzer show a higher abundance of Class II objects (see Sect.~\ref{discuss}). 
We confirm this classification based on the computation of a
single slope ($\alpha$) by applying several alternative methods, 
which are described in Appendix~\ref{appendixA}. 
We also stress that the percentages we estimate are lower limits to the actual number of Class III sources in Lupus~V and VI, 
since our selection criteria require a mid-IR excess and, hence, not all Class III sources are recovered, 
having them in many cases normal stellar photospheric colors. 
Indeed,  X-ray observations of Cr~A revealed a significant larger number of Class III objects in this cloud 
than selected by the Spitzer GB survey (Peterson et al., submitted).

\subsection{Interesting objects}

\subsubsection{Transition objects \label{trans}}

Transition disks represent a critical benchmark for disk evolution
and planet formation models and it is therefore crucial to study
their properties on a statistically significant basis.

\begin{figure*}
\centering
\includegraphics[angle=0,scale=0.6]{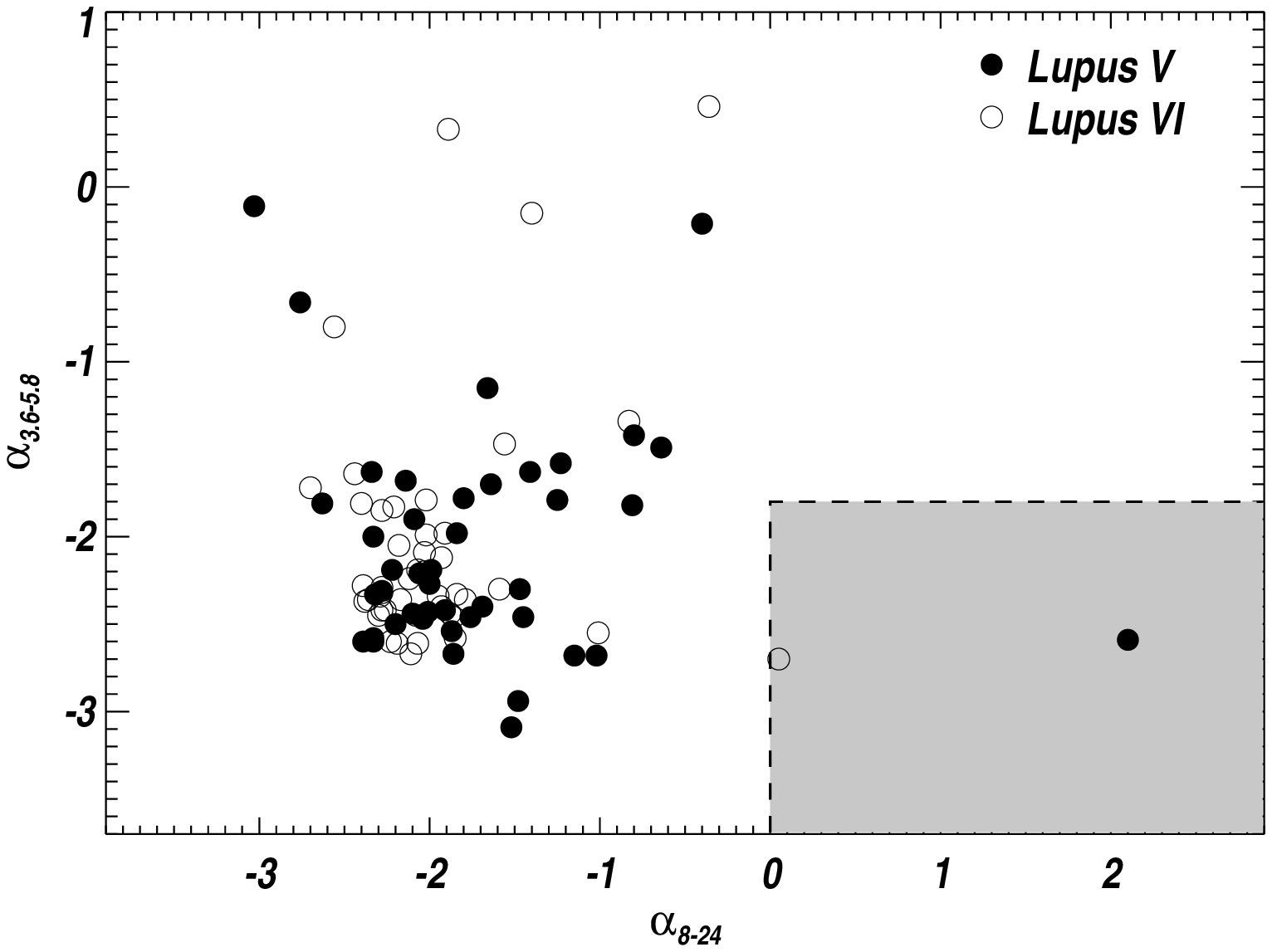}
\caption{SED slope $\alpha_{3.6-5.8}$ as a function of the $\alpha_{8-24}$ slope for YSO candidates 
in Lupus V and VI. The dashed lines define the region where
transition objects are expected ($\alpha_{8-24}>$0 and $\alpha_{3.6-5.8}<$-1.8). \label{alphaalpha}}
\end{figure*}

We searched for possible transition objects among the YSO
candidates identified in both Lupus V and VI. We used the method
proposed by \citet{Muzerolle2010} and plot the SED slope in the 3.6-5.8~$\mu$m
interval ($\alpha_{3.6-5.8}$) versus the SED slope in the 8-24~$\mu$m
interval ($\alpha_{8-24}$). These slope values are reported in
Table~\ref{tab_param_V} and \ref{tab_param_VI}. Figure~\ref{alphaalpha} shows the
$\alpha_{3.6-5.8}$ versus $\alpha_{8-24}$ diagram and the dotted
lines define the region where transition objects are expected
($\alpha_{8-24} >$0 and $\alpha_{3.6-5.8}<$-1.8). We find two 
potential transition objects, namely ID~3 in Lupus V and ID~19 in Lupus VI, which are marked with an asterisk in
Table~\ref{tab_param_V}-\ref{tab_param_VI}. 
The transitional nature of these two objects is also supported on the basis 
of other SED parameters indicating the presence of a inner hole in their disks (see Sect.~\ref{sed_par} and Figure~\ref{alphaturnoff}).

%Based on these results, the frequency of transitional objects in Lupus V and VI is about 2\%, 
%i.e. slightly smaller than measured in other YSO populations identified by Spitzer \citep[4-12\%;][]{Merin2010}.

\subsubsection{Flat sources \label{flat_obj}}

Flat-spectrum sources,  whose emission arises from both a disk and an envelope likely possessing a wind-carved cavity \citep{Greene1994,Evans2009b}, 
are believed to be the boundary between protostars and pre-main sequence objects. 
This phase is one of the most interesting in the evolution of YSOs as phenomena such as outflows and mass accretion,  
which  determine the final mass of the star and process the surrounding material, are particularly active.
 
\begin{figure*}
\centering
\includegraphics[angle=0,scale=0.6]{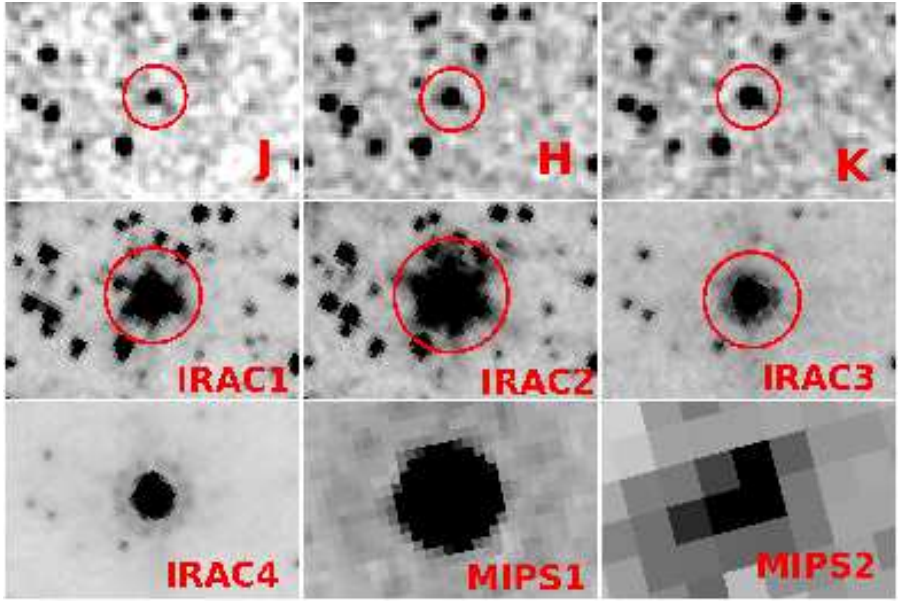}
\caption{2MASS and Spitzer images of the flat-spectrum YSO candidate identified in Lupus~VI (ID~3). 
Each snapshot covers an area of 75$\arcsec \times$50$\arcsec$; north is up and east to the left. \label{flat_sou}}
\end{figure*}

The YSO candidates sample selected in Lupus V and VI contains only one flat sources, namely ID~28 in Lupus VI, and its SED in shown in Figure~\ref{seds_VI}. 
This object in not detected at optical wavelength neither by the NOMAD nor by the DENIS surveys, as expected for young objects still partially embedded, and 
appears point-like in all the 2MASS and Spitzer 3.6-24~$\mu$m images (Figure~\ref{flat_sou}). 
This source was recently observed, and not detected, in the HCO$^+$(3-2) sub-millimeter line by Amanda Heiderman (private communication) 
using the 10m Caltech Submillimeter Observatory (CSO) in Mauna Kea (Hawaii). 
\citet{Kempen2009} presented a YSO classification method based on HCO$+$ lines that successfully separates embedded 
Class I sources (strong HCO$+$ emission) from edge-on Class II disks and confused sources (little or no HCO$+$ emission). 
The lack of detection in the HCO$^+$(3-2) line casts doubt on the existence of a residual dense envelope around our flat-spectrum YSO candidate. 
The upper limit to its main beam HCO$+$ line temperature, computed as 2$\sigma_{rms}$, is $<$0.48~K.

\subsection{Spatial distribution and clustering of the YSO candidates \label{clustering}}

The spatial distribution of YSOs in relation to the cloud structure can be used to study a very specific  aspect of the star formation process, 
i.e. whether it occurs in a centrally condensed way  or there is more than one center of density  \citep[see][ and references therein]{Lada2003}. 

\begin{figure*}
\centering
\includegraphics[angle=0,scale=0.8]{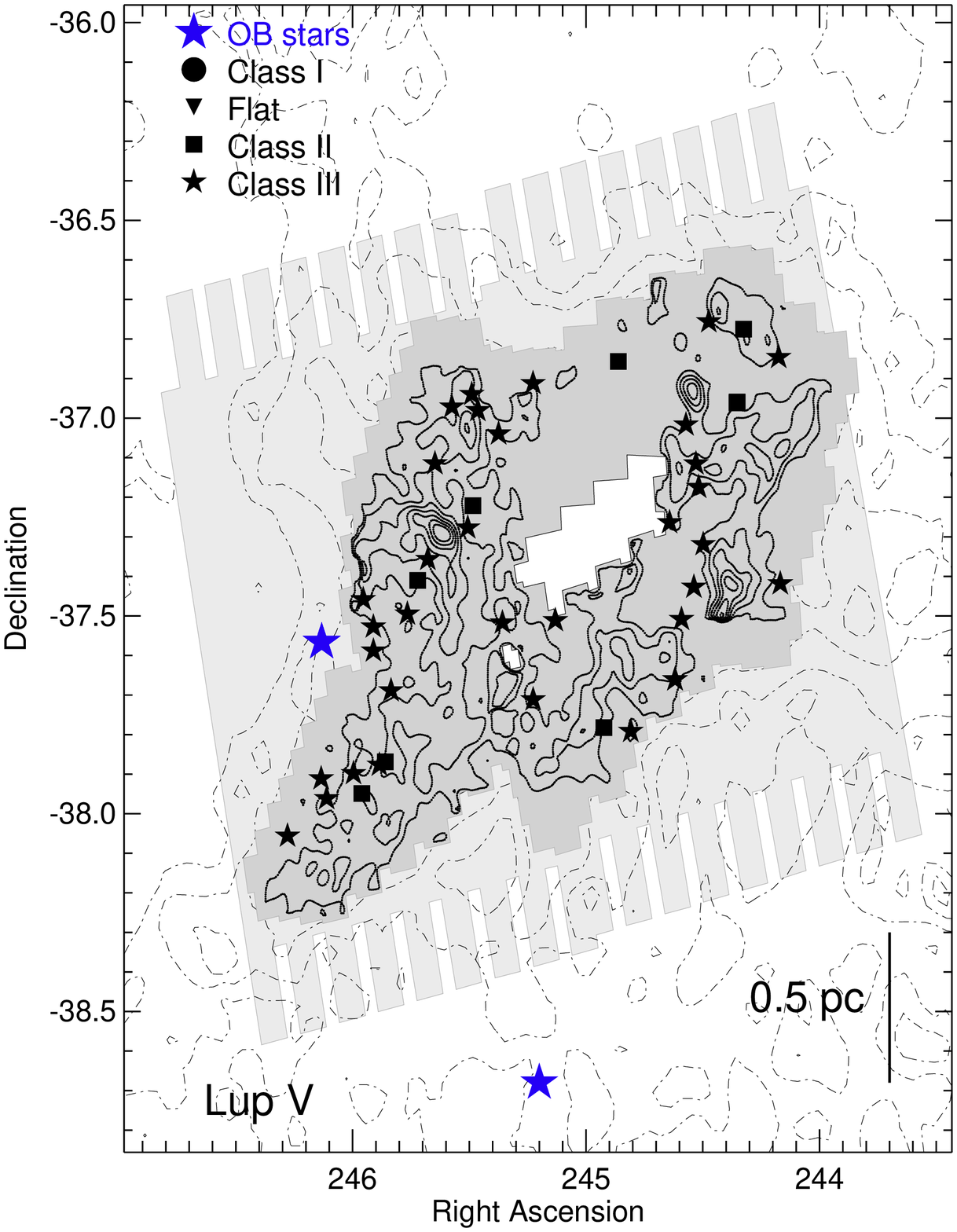}
\caption{Visual extinction (A$_V$) contours of the Lupus~V cloud, from 2 to 20 mag in steps of 2 mag, from the Spitzer GB extinction
map (solid lines). The intersection of the areas observed with the four IRAC bands (dark grey) and the area observed with 
the MIPS~1 band (light grey) are also shown. The dashed lines outside the
IRAC area are the contour levels of extinction from \citet{Cambresy1999}, from 1 to 6 mag in steps of 0.35 mag. 
Symbols represent the location of the YSO candidates in this cloud as
a function of the Lada class, as explained in the legend. The position of the OB stars in the field is also indicated. \label{OB1}}
\end{figure*}

\begin{figure*}
\centering
\includegraphics[angle=0,scale=0.7]{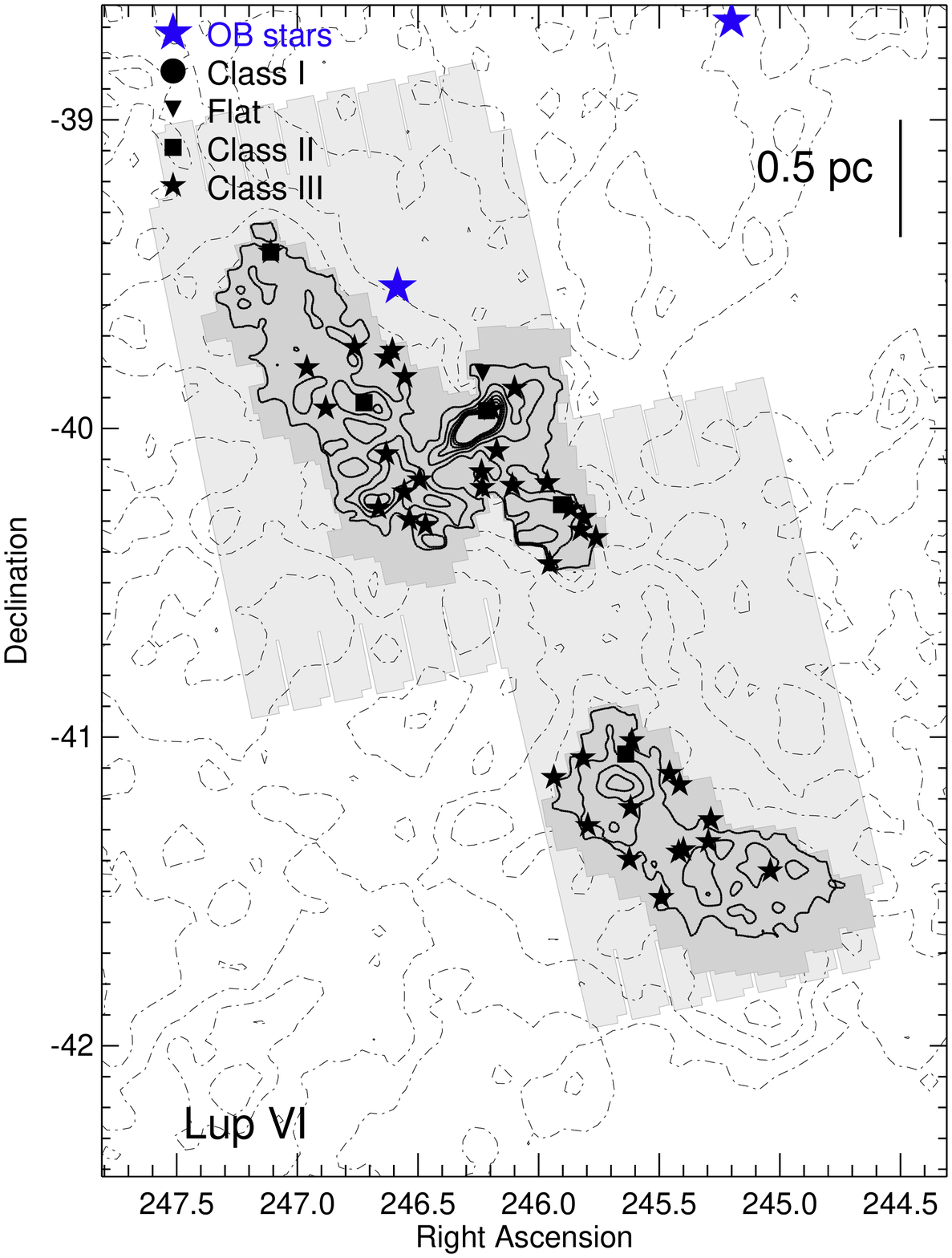}
\caption{Same as Figure~\ref{OB1} for Lupus VI. \label{OB2}}
\end{figure*}

Figures~\ref{OB1} and \ref{OB2} show the spatial distribution of the YSO candidates in the Lupus V and VI clouds on the relative extinction maps 
obtained from the Spitzer GB data. These extinction maps have a spatial resolution of 120~arcsec and the details 
on how such maps were derived are given in the final c2d data delivery document \citep{Evans2007}. 
In both clouds,  the regions of high extinction (A$_V \gtrsim$6~mag) are rare. 
The great majority of the YSO candidates (of any SED class) lie in
the vicinity of the highest extinction regions, however their position does not correlate with the location of the extinction peaks. 
A possible explanation is that star formation may have not taken place in these clouds very recently (i.e. more than  5-10 Myrs ago), 
so that their YSO populations appear now to be dispersed in the 
surroundings of the extinction peaks. This interpretation would fit with the high abundance of Class III 
objects in both clouds, but needs to be confirmed by velocity dispersion and age measurements (see Sect.~\ref{discuss}).

\begin{figure*}
\centering
\includegraphics[angle=0,scale=0.9]{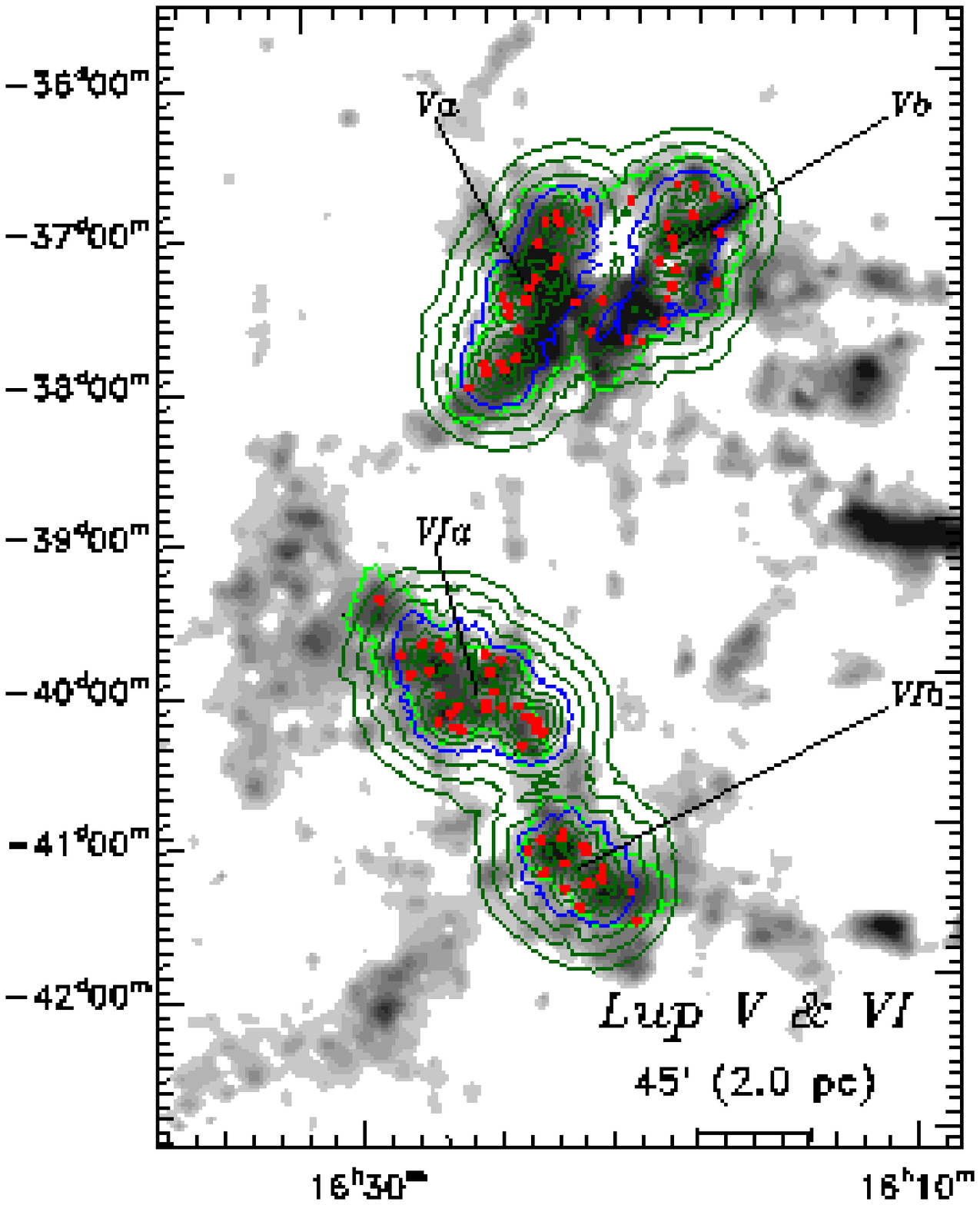}
\caption{Volume density plot for YSO candidates in Lupus V and VI as determined with the nearest-neighbor algorithm. 
The derived density contours are compared with the GB extinction map (in gray scale). The blue contour corresponds to the 
1.0~M$_\odot / pc^3$ density level, while the green ones correspond to levels of 0.125, 0.25, 0.5, 2, 4, and 8 times this level. 
The four "loose groups" are identified with labels. Red filled circles represent the YSO candidates. 
The thicker green polygons represent the areas observed with both IRAC and MIPS.  \label{clust_fig}}
\end{figure*}

To investigate quantitatively the clustering properties of the YSOs candidates in Lupus V and VI, 
we applied \citet{Casertano1985} surface/volume density algorithm, 
following the same approach as for the c2d clouds \citep[see, e.g.,][]{Evans2009a}. 
In particular, the criterion we adopt to identify YSO structures is described in detail by \citet{Jorgensen2008} and divides concentrations of YSOs into ``clusters'' or ``groups''. 
``Clusters'' are defined as regions with more than 35 YSOs within a given volume density level and "groups" as regions with less. 
Clusters and groups can be ``loose'', if their volume density is higher than 1~M$_\odot / pc^3$, or  "tight" if their volume density is higher than 25~M$_\odot / pc^3$. 
Figure~\ref{clust_fig} shows the result of the clustering analysis for Lupus V and VI. 
The YSO density contours are plotted, for comparison, on the extinction map. The YSO candidates are divided up into 4 loose groups, two in each cloud. 
While the two groups observed in Lupus V are definitively real, the discontinuous coverage of the IRAC 
observations in Lupus VI might have biased the algorithm result toward the identification of two separate YSO groups in this cloud. 
Furthermore, there are some low A$_V$ clumps in the eastern part of the region with no associated YSO candidates; this is simply due to the fact that our
Spitzer observations did not cover this part of the sky. 
Table~\ref{clust_tab} summarizes the results of the clustering analysis in Lupus V and VI; we report the number 
of YSO candidates, their Lada class, the cloud mass, the volume and the star formation efficiency (SFE) measured in each of the four loose groups. 
The cloud mass (M$_{cloud}$) was determined from the GB extinction map, while the SFE is defined as $M_{stars}/(M_{stars}+M_{cloud})$, 
where the total mass converted into stars (M$_{stars}$) was estimated assuming an average YSO mass of 0.5~M$_\odot$, 
to be consistent with the estimates in Table~12 by \citet{Merin2008} for the other Lupus clouds, which we use for comparison. However, as we will see in Sect.~\ref{LF}, 
the YSO population in Lupus V and VI might have a characteristic stellar mass  between 0.6 and 1.3~$M_{\odot}$ depending on the assumed age, higher 
than the other Lupus cloud and slightly higher than the peak of the standard IMF \citep[e.g.,][]{Miller1979,Chabrier2001,Kroupa2005}. 
Thus, in Table~\ref{clust_tab} we also include SFE calculations assuming the standard form of the IMF by \citet{Kroupa2005}.

Lupus V and VI clearly show multiple density structures with separate gas concentrations that evolve independently and a SFE around 3-4\% 
(assuming 0.5~M$_\odot$ as the typical YSO mass), similar to the SFE measured by \citet{Merin2008} for Lupus I and IV but lower 
than the SFE measured for Lupus III  (see their Table~12). 
Overall, both the cloud density structure and the distribution of YSO candidates in the two regions suggest a 
dispersed distribution of volume density enhancements  rather than a centrally-condensed 
structure. This dispersed distribution of YSOs, with peak densities of 2-4 YSO candidates per pc$^{3}$ following the lane of the dust emission, is typical of 
Lupus I and IV, while in the more active Lupus III a centrally-condensed structure appears to 
dominate the star-formation process \citep{Merin2008}. 
In particular, the clustering analysis shows that the spatial distribution of Class III objects in Lupus V and VI is similar to that of Class III members in Lupus I and IV, 
while it is clearly more spread than the Class III population in Lupus III. This larger spread might indicate that the Lupus I and IV population is  a few Myrs older than 
the Lupus~III population, but  again this hypothesis needs to be confirmed by velocity dispersion and age measurements. 
As we will see Sect.~\ref{discuss}, a few Myrs older age (5-10~Myrs) for the stellar population in Lupus V and VI with respect to Lupus III 
would explain both the higher spatial dispersion and the higher fraction of Class III objects in these two clouds. 
Considering this age and given that the YSO candidates in Lupus V and VI are approximately dispersed over a 4~pc radius in both clouds (Figure~\ref{clust_fig}), 
we expect their velocity dispersion to be of the order of 0.4-0.8~km/sec, which well matches the range observed 
for other galactic star forming regions and young clusters \citep[see, e.g.,][and reference therein]{Kraus2008}.

\section{Discussion \label{discuss}}

In the previous section we have shown that most YSO candidates in Lupus V (79$\%$)
and Lupus VI (87$\%$) are surrounded by optically thin disks (Class III objects). 
To place our results in a more global context, we report in Table~\ref{classes} 
the number of YSO candidates organized by IR class for all the Lupus clouds 
observed with Spitzer \citep[Lupus I, III, IV, V, and VI; this work and][] {Merin2008} and the
other clouds observed with Spitzer within the frame of the c2d survey, namely Cha~II \citep{Alcala2008}, Perseus (Lai et al., in preparation),
Serpens \citep{Harvey2007b}, and Ophiuchus (Allen et al., in preparation). 
We find a much higher number of the Class III objects in both Lupus V and Lupus VI than in Lupus I, III and IV. 
The difference gets even stronger when compared to the fraction of Class III objects measured in the c2d cloud sample and summarized in \citet{Evans2009a}. 
To explain the origin of the surprisingly high fraction of Class III objects in Lupus V and Lupus VI, 
we considered the following possible scenarios and investigate them in more details in Sect.~\ref{sect_contam}-\ref{LF}:

\begin{enumerate}

\item {\bf Contamination:} Contamination by field objects with small IR excess emission 
can affect the measured fraction of Class III YSO candidates.

\item {\bf Binarity:} Most circum-binary disks are cleared by ages of 1-2 Myr, 
while most circumstellar disks are not \citep{Bouwman2006}. 
Thus, a significantly higher fraction of binary stars in Lupus V and VI with respect to the other Lupus clouds 
would explain the higher fraction of Class III sources. 
 
 \item {\bf Stellar age:} Because Class II objects are very common at ages $\sim$1~Myr and very rare at $\sim$10~Myr (Sect.~\ref{intro}), 
 a few Myrs older age for the stellar population in Lupus V and VI with respect 
 to the other Lupus clouds \citep[1.5-4 Myr;][]{Hughes1994, Comeron2003} would also explain the higher fraction of Class III objects.

\item {\bf Disk photo-evaporation:} This scenario would rely on the presence of OB stars lying close enough to 
the Lupus V and VI YSO populations to photoevaporate their disks, but too far from the Lupus I, III and IV populations
to have any influence on the evolution of their disks.

\item {\bf Characteristic stellar mass:} Recent studies suggest a much shorter disk
dissipation time \citep{Brown2007,Kim2009,Merin2010} and higher mass accretion rate for higher mass stars 
\citep{Natta2004,Natta2006,Sicilia-Aguilar2006}. Thus, a
possible difference between the characteristic stellar mass in
Lupus V and VI and the other Lupus clouds may be the cause of the
different evolutionary stages of the observed disks.

\end{enumerate}

Using our dataset and complementary information from the literature we now investigate these scenarios.

\subsection{Contamination \label{sect_contam}}

The immediate explanation would be that our YSO candidates sample is significantly contaminated by field objects with small IR excess emission similar to that  of Class III YSOs. 
The adopted YSO selection method has proven to be successful since most YSOs selected in the c2d clouds have been
confirmed via spectroscopy \citep{Spezzi2008,Merin2008,Oliveira2009,Cieza2010}. The removal of extra-galactic objects is optimal 
and the fraction of contaminants, mainly consisting of AGB stars with small excess emission in IRAC and/or MIPS~24$\mu$m mimicking the behavior of Class III YSOs, 
is estimated to be around 30\% \citep[e.g.][]{Oliveira2009,Cieza2010}. 
This estimate is based on the spectroscopic follow-up of Class III objects in the Serpens molecular cloud, which is located at b$\approx +$5~deg from the galactic plane. 
Thus, we expect the higher galactic latitude of Lupus V and VI (6$< b <$8~deg) to minimize this problem. 
However, even taking a 30\% contamination level into account, the fraction of Class III YSO candidates 
in Lupus  V and IV would reduce to 70\% and 81\%, respectively, and is still much higher than measured in other c2d clouds.

To verify further the amount of expected contamination, we have also estimated 
the numbers of background AGB stars expected towards the direction of Lupus V and VI over the areas observed with Spitzer. 
We performed this exercise by using the Galaxy model by \citet{Robin2003} and their online tool\footnote{http://model.obs-besancon.fr/}. 
According to this model, we expect less than 1 background AGB stars in the 3.82 and 2.88~deg$^2$  areas observed in Lupus V and VI, respectively, 
with apparent $J$ magnitude between 5 and 16.5, i.e. the photometric ranges covered by our YSO candidates (Tables~\ref{tab_mag_LupV} and \ref{tab_mag_LupVI}). 
Galactic counts models become more and more inaccurate towards the very low-mass regime and the predicted number depends on the adopted model, however this 
exercise ensures that AGB stars can not contribute noticeably to the contamination of our Class III YSO candidates.

\subsection{Binarity \label{binary}}
 
It has been shown that interactions with close
stellar or planetary companions can significantly influence the
evolution and lifetime of protoplanetary disks \citep{Monin2007}.
In particular, \citet{Kraus2010} have shown that disk evolution of close (5-30~AU) binary systems is very
different from that of single stars, and most circum-binary disks are
cleared by ages of 1-2 Myrs, while most circumstellar disks around single stars are not \citep{Bouwman2006}.  
Thus, a significantly higher fraction of binary stars in Lupus V and VI with respect to the other Lupus cloud 
would also explain the high fraction of Class III sources.  
Specifically, the fraction of binary stars in Lupus V and VI should be about 50\% higher than in the other Lupus clouds 
to account for the relative abundance of Class III YSOs.

There are no indications in the literature of a higher binary fraction in Lupus V and VI  
relative to other Lupus clouds and, hence, this scenario appears unlikely. 
However, we cannot definitively rule it out. High resolution spectroscopy \citep[see, e.g.,][]{Prato2007} 
and/or adaptive optics and long-baseline interferometry would be
necessary to properly measure the binary fraction in these clouds \citep[e.g.,][]{Raghavan2007}.

\subsection{Stellar age \label{age}}

IR studies of individual star-forming regions have long suggested that disk lifetimes are relatively short 
\citep[$\lesssim$5~Myrs;][]{Strom1989,Skrutskie1990,Lada1995,Brandner2000}. 
\cite{Haisch2001} were the first to provide more precise constraints. They showed that the fraction of stars with disk
within a cluster rapidly decreases with age, such that one-half the stars lose their disks within the first 3~Myr, and derived an overall disk lifetime of $\sim$6~Myr. 
More recently, \citet{Evans2009a} derived half-life for each of the Lada classes from the combined analysis of the Spitzer c2d dataset. 
According to this study, the half-life for Class II sources is $\sim$2 Myrs. Thus, an age of $\sim$10~Myr for the young population in Lupus V and VI, i.e. only a few Myrs 
older than the other Lupus clouds \citep[1.5-4 Myr;][]{Hughes1994, Comeron2003}, 
would explain the higher fraction of Class III objects in these clouds.

We can not assess the presence of age differences of the order of a few Myrs among the Lupus clouds on the basis of the 
Spitzer and complementary data available at the moment in the literature. 
Optical photometry and spectroscopy would help to estimate isochronal ages and/or  ages based on the  LiI~670.8nm line for the YSO population in each Lupus cloud. 
However, these methods provide only weak constraints on the age \citep[see, i.e.,][]{Palla2005}, so that 
a possible age difference of a few Myrs  between the Lupus V and VI population and 
the population in other Lupus clouds would not be clearly reflected by their LiI~670.8nm absorption line or position in optical/near-IR color magnitude diagrams. 
Moreover, the reliability of lithium abundance as an age indicator is even more limited after the 
recent finding that events of episodic accretion enhance the lithium depletion of low-mass stars and brown dwarfs \citep{Baraffe2010}.
As for the isochronal age, the uncertainty on the distance to the Lupus clouds prevents a reliable derivation 
of the age based on the position of the objects in the temperature-luminosity diagram. 
This problem will be solved in the near future thanks to the forthcoming GAIA mission.

Beside the impossibility to put accurate constraints to the age of Lupus V and VI  on the basis of Spitzer data, 
it is plausible that in these two clouds star formation ceased almost entirely a few Myrs ago while this is not the case for other Lupus clouds. 
\citet{Heiderman2010} investigate the relation between star formation rate (SFR) and gas surface densities ($\Sigma _{gas}$) for 
20 large molecular clouds from the c2d and GB surveys. 
This study revealed that the SFR surface density ($\Sigma _{\rm SFR_{\rm IR}}$) versus $\Sigma _{gas}$ relation 
is a power law whose slope changes from a steep relation at $\Sigma _{gas}< \Sigma _{th}$ 
(slope of $\sim$4.6) to a linear relation (slope of $\sim$1.1) above $\Sigma _{th}$. 
The authors estimate $\Sigma _{th}=$129$\pm$14~M$_\odot /pc^{2}$, corresponding to A$_V \approx$8.6~mag, and denote this as a star formation threshold. 
This does not imply no star formation below the threshold, but clearly indicates that the probability to find star forming cores is higher above the  threshold.
The star-forming threshold found by \citet{Heiderman2010} is in agreement with the threshold (A$_V \approx$7-10~mag) found in other studies of local molecular clouds 
\citep{Onishi1998,Johnstone2004,Enoch2007,Lada2010,Andre2010}. 
In other words, if most of the present-day mass measured for a given cloud lies below A$_V \approx$8~mag, 
a decrease in star formation could plausibly be caused by exhaustion of gas above such a threshold in surface density. 
In Table~\ref{mass_tab} we report the fraction of cloud mass below the threshold for Lupus V and VI and compare it with the fraction measured for the other Lupus clouds, 
in particular for Lupus~III which is the most active star forming cloud in the complex  and is mainly populated by Class~II objects (see Table~\ref{classes}). 
It is clear that the cloud mass above threshold in Lupus V and VI
is very small or zero, while in Lupus~III and in the other Lupus clouds a significant fraction of the cloud mass is still above threshold. 
Thus, it is not surprising that star formation is still ongoing in Lupus~III, as reflected by the substantial number of objects 
in younger SED classes. In contrast, the absence of such sources in Lupus V and VI could be explained 
by a cessation of active star formation a few Myrs ago. 
It may be hard to get an actual age difference between Lupus V and VI on the one hand and the other Lupus clouds on the other, both for the reasons mentioned above and because 
the oldest stars in each cloud may have similar ages. Table~\ref{mass_tab} only indicates that 
the duration of the star-forming event in Lupus V and VI was less because it started with less dense gas and used it up longer ago.

\begin{figure*}
\includegraphics[angle=0,scale=0.6]{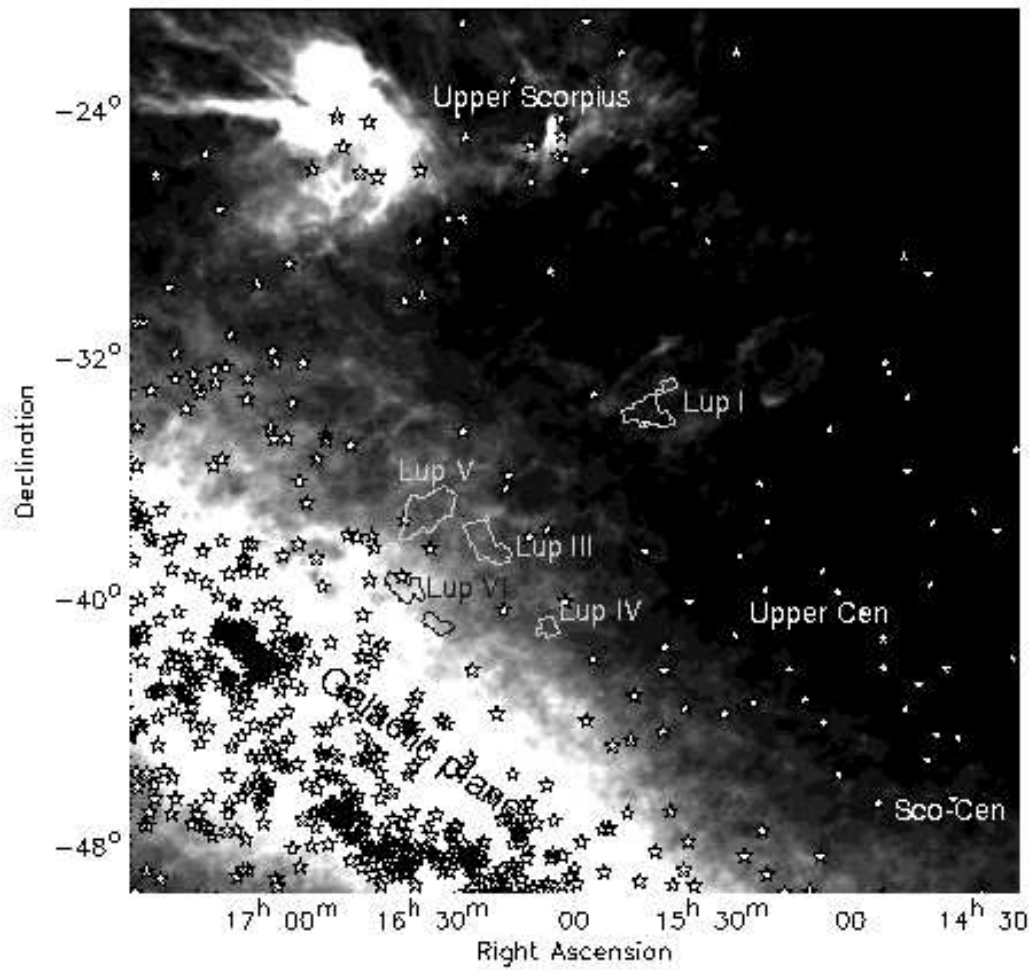}
\caption{IRAS~100$\mu$m dust emission map of the Lupus cloud complex from the Galactic Dust Reddening and Extinction database at NASA/IPAC 
Infrared Archive. The polygons represent the areas mapped with IRAC and MIPS 
in Lup I, III, IV, V and V as part of the c2d and GB surveys. The stars indicate the position of the OB stars from the catalog by C. Reed (private communication). 
The positions of the Galactic plane and the  Scorpius-Centaurus main OB associations are also indicated. \label{fig_OB}}
\end{figure*}

Another argument in favor of an older age of Lupus V and VI with respect to other Lupus clouds comes from their spatial location with respect to the 
Sco-Cen association. Lupus I, III, IV are all at high galactic latitude($b \approx$+9, +16 and +48, respectively), in the region of the Lupus complex closest to Sco-Cen 
\citep[see Figure~\ref{fig_OB} and Figure~2 by][]{Tachihara2001}. 
In particular, Lupus~I is at the edge of an HI shell left over from a supernova event in the Upper-Sco 
sub-group of Sco-Cen about 1.5 Myr ago \citep{Tachihara2001}. 
The dynamics of the HI and the CO towards Lupus I are consistent with 
the molecular cloud being compressed by the HI shell \citep{Tothill2009}, and this may have caused recent/ongoing star 
formation. Lupus~III and IV lie in ``streamers'' that stretch from the main body of 
Lupus towards the Upper-Cen-Lup sub-group of Sco-Cen, but may also be 
exposed to effects from Upper-Sco. Thus, even though Lupus III and IV do not fit the argument so well as Lupus~I, 
recent triggered star-formation might play a role in their star formation history, whereas Lupus V and VI lie at lower galactic latitudes ($b \approx$+8 and +6, respectively) 
and do not show any morphological evidence of interaction with Sco-Cen.

\subsection{OB stars and disk photoevaporation \label{OBstars}}

Circumstellar disks in stellar aggregates can be effectively evaporated within a few Myrs through the action of external ultraviolet radiation from the most massive stars in the group. 
The intensity and spatial extent of such radiation fields depends on the number and spectral type of the massive stars.
For small groups of 100/500 members, which is the case of the Lupus complex, photoevaporation of disks around low-mass stars (M$\lesssim$1~M$_\odot$) 
is expected to be efficient within about 0.5-1~pc of the massive stars \citep[see][and references therein]{Johnstone1998,Storzer1999,Adams2004,Balog2007,Guarcello2009}. 
Indeed, it has been recently shown that in the young open
clusters NGC~2244 and NGC~6611, both containing a rich low-mass PMS
population as well as a large number of OB stars, the
disk frequency drops within a distance of $\sim$0.5~pc
from the massive star group \citep{Balog2007,Guarcello2009}. At larger distances, however, stars with
disks do not appear affected by the presence of massive stars, since the
disk frequency is not spatially correlated with the positions of massive stars.
We can consequently assume that any disk photoevaporation effect is expected to be seen within $\sim$1~pc from the ionizing OB stars.

We used the OB star catalog provided by C. Reed \citep{Reed2003} and searched for OB stars lying in the vicinity, i.e. within 1~pc, of
the Lupus clouds. Figure~\ref{fig_OB} shows the spatial
distribution of massive OB stars  together with the areas observed by Spitzer in Lupus I, III, IV, V and V as part of the c2d and GB surveys. 
In Table~\ref{OB_tab} we report the number of OB stars within 1~pc of each of the Lupus clouds, their spectral type and luminosity and the projected distance 
from the cloud.  Because of the paucity of OB stars in the vicinity each cloud, 
it is very unlikely that photoevaporation plays a relevant role in the disk evolution of the YSOs in these clouds. 
In particular, for Lupus V and VI we found just one star lying within 1~pc from each cloud, as shown in more detail in Figure~\ref{OB1} and Figure~\ref{OB2}, and this 
number is comparable to that found in Lupus I, III and IV. Thus, there is no evidence that disk photoevaporation is acting more efficiently in Lupus V and VI then in the other Lupus clouds. 
Moreover, both for Lupus V and VI, we did not find any spatial correlation between the Class II disk frequencies and the position of these
massive stars. For instance, in Lupus VI, Class II YSO candidates are more abundant within 1~pc from the only B-type star in the field than further away. 
Another argument discharging the photoevaporation hypothesis is that the radius up to which a star can externally photoevaporate a disk depends on the square root 
of its ionizing flux, which in turn depends on the spectral type \citep[see, e.g.,][]{Johnstone1998,Storzer1999,Adams2004,Balog2007,Guarcello2009}. 
Radii of the order of 0.5-1~pc apply to early and mid-type O stars, 
which have ionizing fluxes orders of magnitude greater than those of mid or late B stars. As Table~\ref{OB_tab} indicates, 
the nearest stars to the Lupus clouds have mid or late B types, and are thus virtually harmless to any nearby disks. 

For these overall reasons, it is very unlikely that the proximity of massive stars could be the cause of the high
abundance of Class III YSO candidates in Lupus V and VI.

\subsection{Luminosity function and characteristic stellar mass \label{LF}}

\begin{figure*}
\includegraphics[angle=0,scale=0.9]{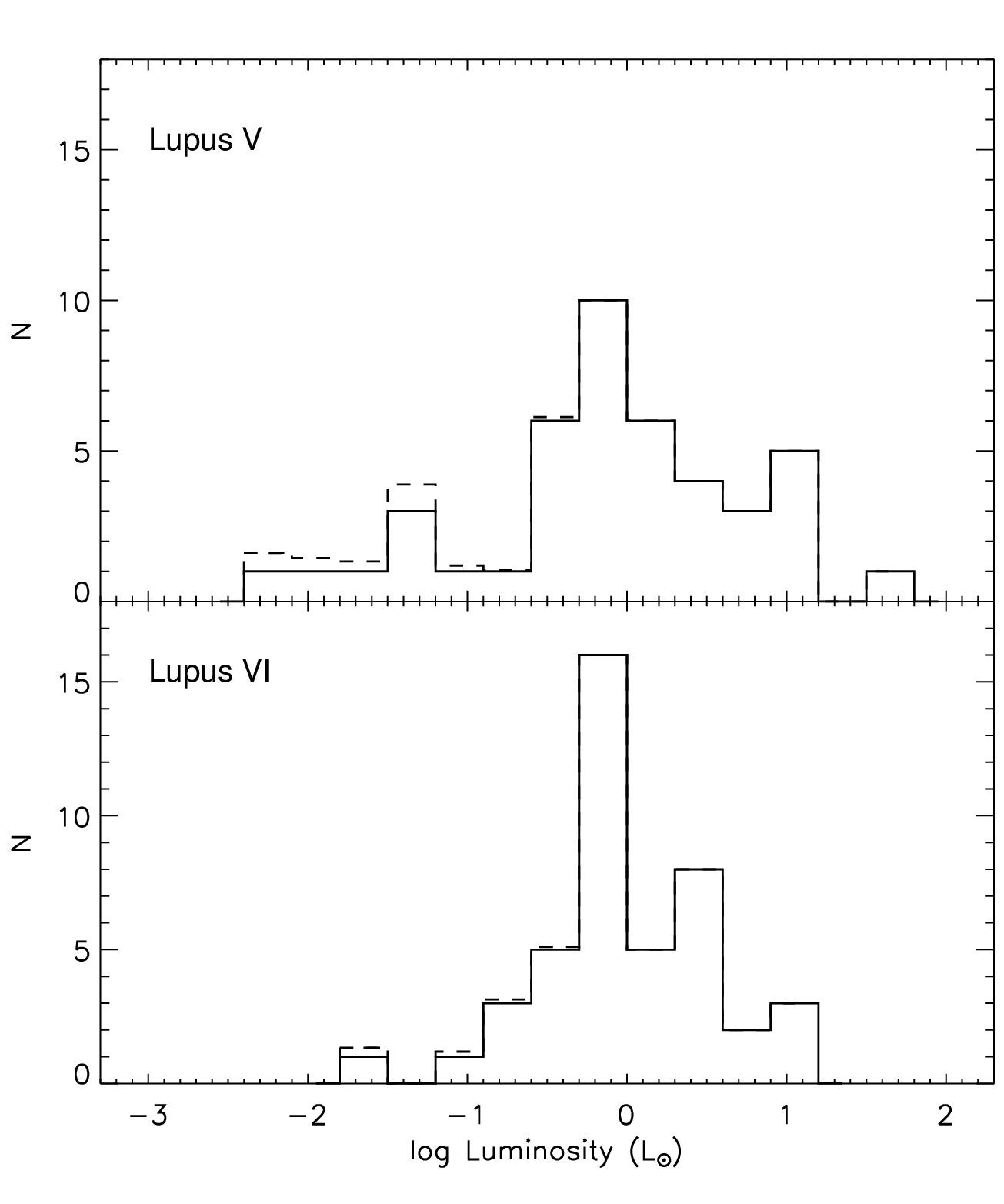}
\caption{Bolometric luminosity function for the YSO candidates in Lupus V and VI before (solid line) and after correction for completeness effects (dashed line). \label{lumitot}}
\end{figure*}

The determination of the stellar luminosity can be used as a poor but still useful first 
order proxy for mass, assuming that most of the stars have formed more or less at the same time. 
Before using the range of luminosities to provide an estimate of the mass range, we determined the degree of completeness of the 
YSO candidate sample in Lupus V and VI. 
We used the same approach as \citet{Merin2008} and derive first the bolometric luminosity function for Lupus V and VI. 
The bolometric luminosity of each YSO candidate was obtained by integrating
over the entire observed SED flux; the total flux was finally
converted to luminosity assuming a distance of 150~pc for both Lupus V and VI clouds \citep{Comeron2009}. 
We then applied \citet{Harvey2007b}'s completeness estimate for the c2d survey to our 
YSO candidate samples, since the GB catalogs have the same photometric
depth as the c2d ones. \citet{Harvey2007b} made an estimate of the
completeness of the c2d catalogs by comparing, for each luminosity
bin, the number of counts from a trimmed version of the deeper SWIRE catalog of extragalactic
sources (assumed to represent 100$\%$ completeness by c2d
standards) with the number of counts for the c2d catalogs in
Serpens. This completeness estimate has been applied to all five
molecular clouds observed within the c2d frame, under the
assumption that the Spitzer observations for these clouds were
homogeneous in terms of photometric depth. 
Figure~\ref{lumitot} shows the bolometric
luminosity function  for Lupus V and VI before (solid line) and after (dashed line) correction for completeness and
suggests that we are missing only a few low-luminosity objects with $L<0.1 L_{\odot}$, 
i.e. $\sim$4 in Lupus V and $\sim$1 in
Lupus VI. These objects have been missed by our selection either because they are below
the noise level of our GB observations or because they are located
within the galaxy loci of the CM diagrams (Figure~\ref{LupV_sel}-\ref{LupV_sel}). 
In conclusion, both luminosity histograms suggest a 
completeness better than $\sim$85\% at luminosities down to 0.1~L$_{\odot}$ 
for the YSO candidate samples in Lupus V and VI, which corresponds to
a mass of 0.1, 0.2 and 0.35~M$_\odot$ for 1, 5 and 10~Myrs old stars, respectively, according
to the PMS evolutionary tracks by \citet{Baraffe1998}.

Both Lupus V and VI clouds the peak of the luminosity
function appears between 0.5 and 0.6~L$_\odot$, which corresponds
to a 0.6~M$_\odot$ star at an age of 1 Myr and to a 1.3~M$_\odot$ star at an age of 5 Myr. 
Note that most young stars within the Lupus complex have an estimated age which falls within the 1-5~Myr
interval \citep{Comeron2008_rev}. The bolometric luminosity histogram for the other Lupus clouds
\citep[see Figure~13 by][]{Merin2008} suggests a typical
luminosity of 0.2~$L_{\odot}$, which corresponds to a 0.2~M$_\odot$ star at an age of 1 Myr and to a 1~M$_\odot$ star at an age of 5~Myrs. 
Thus, YSO candidates in Lupus V and VI are on the average
 0.4~M$_\odot$ more massive than YSOs in the other Lupus
clouds. As mentioned in Sect.~\ref{discuss}, this difference may
have some impact on the observed properties of disks because the
disk lifetime and mass accretion rate both appear to depend on stellar
mass \citep{Brown2007,Kim2009,Merin2010,Muzerolle2010,Sicilia-Aguilar2010}. 
Because of the large uncertainties and limitations affecting
the measurement of both the disk lifetime and the mass accretion
rate, it is still very premature to give a disk lifetime versus
stellar mass calibration relation and, hence, estimate whether a
0.4~M$_\odot$ difference in the characteristic stellar mass is
significant. Moreover, the Lupus complex extends over about 50~pc along the line of sight \citep[see Sect.~3 by][]{Comeron2009}.  
Thus, the difference in the characteristic stellar mass we measure among the Lupus clouds might 
be partly due to the use of inaccurate distances for each individual cloud. 
However, the incorrect distance can hardly be the only cause of the greater peak in the luminosity function of Lupus V and VI, 
since bringing down the peak in luminosity (and therefore in mass) by a factor of $\sim$3 would require 
these clouds to be at $\sim$70\% of the assumed distance, placing them at 100~pc or less from the Sun, which seems unlikely. 

Thus, it is worth mentioning that the higher
characteristic stellar mass of Lupus V and VI might be a contributing
factor to the high fraction of Class III YSO candidates observed in these clouds.

%%%%%%%%%%%%%%%  Tables

\begin{table*}
\caption{Lupus V and VI detection statistics. \label{detection_stat}}
\begin{tabular}{lll}\hline\hline
Detection criterion                               & ~~~~~~~~~~~Number of sources      \\\hline
                                                                & Lupus V   & Lupus VI  \\\hline
IRAC~3.6$\mu$m with S/N$\ge$5                       & 292611    & 243185      \\
IRAC~4.5$\mu$m with S/N$\ge$5                       & 179095    & 168863    \\
IRAC~5.8$\mu$m with S/N$\ge$5                       & 33870     & 32436 \\
IRAC~8.0$\mu$m with S/N$\ge$5                       & 18366     & 17548 \\
All 4 IRAC bands with S/N$\ge$5                     & 14571     & 13632    \\
MIPS~24$\mu$m with S/N$\ge$5                        & 1816      & 2469      \\
MIPS~70$\mu$m with S/N$\ge$5                        & 126       & 87      \\
All 4 IRAC bands, MIPS~24 $\mu$m, \& K with S/N$\ge$5           & 558       & 526
\\\hline
\end{tabular}
\end{table*}

\begin{table*}
\begin{center}
\begin{small}
\caption{IRAC and MIPS observed fluxes for the YSO candidates in Lupus V. \label{flux_LupV}}
\begin{tabular}{llllllll}
\tableline
\tableline
  ID & Name/Position   &  3.6 $\mu$m    &    4.5 $\mu$m      &   5.8 $\mu$m       &   8.0 $\mu$m     &    24 $\mu$m     &     70 $\mu$m     \\
     & (SSTgbsJ)       &  (mJy)         &    (mJy)           &    (mJy)           &    (mJy)         &       (mJy)      &    (mJy)         \\\hline
  1    & 16163197-3704563  &  2000.00$\pm$169.00 & 2270.00$\pm$132.00 & 2360.00$\pm$113.00 & 1790.00$\pm$ 90.80 & 259.00$\pm$26.80 &  81.70$\pm$10.20\\
  2    & 16164035-3725054  &  1860.00$\pm$109.00 & 1500.00$\pm$ 80.90 & 1330.00$\pm$ 64.70 & 1010.00$\pm$ 48.20 & 499.00$\pm$46.60 &	    --       \\
  3    & 16164198-3650456  & 	 1.54$\pm$  0.08 &    1.03$\pm$  0.05 &    0.72$\pm$  0.05 &	0.41$\pm$  0.04 &  12.20$\pm$ 1.17 &	    --       \\
  4    & 16171811-3646307  & 	 4.64$\pm$  0.24 &    3.96$\pm$  0.19 &    3.67$\pm$  0.18 &	3.89$\pm$  0.20 &   5.78$\pm$ 0.58 &	    --       \\
  5    & 16172474-3657333  &   178.00$\pm$  8.60 &  167.00$\pm$  8.42 &  146.00$\pm$  6.91 &  122.00$\pm$  5.91 & 152.00$\pm$14.10 & 468.00$\pm$49.20\\
  6    & 16172485-3657407  & 	21.90$\pm$  1.04 &   14.30$\pm$  0.70 &    9.79$\pm$  0.47 &	5.58$\pm$  0.28 &   4.74$\pm$ 1.17 &	    --       \\
  7    & 16175383-3645207  &  2280.00$\pm$293.00 & 3200.00$\pm$221.00 & 3500.00$\pm$167.00 & 2200.00$\pm$119.00 & 237.00$\pm$22.00 &	    --       \\
  8    & 16175958-3719083  &   205.00$\pm$ 10.10 &  120.00$\pm$  6.02 &  104.00$\pm$  4.91 &   72.40$\pm$  3.52 &  33.90$\pm$ 3.16 &	    --       \\
  9    & 16180425-3710277  & 	21.20$\pm$  1.03 &   13.20$\pm$  0.63 &    9.84$\pm$  0.48 &	7.24$\pm$  0.34 &   1.58$\pm$ 0.25 &	    --       \\
 10    & 16180702-3706544  &  2460.00$\pm$144.00 & 1780.00$\pm$123.00 & 1590.00$\pm$ 78.90 & 1090.00$\pm$ 52.50 & 328.00$\pm$30.60 &	    --       \\
 11    & 16180915-3725364  &   248.00$\pm$ 12.70 &  163.00$\pm$ 10.00 &  132.00$\pm$  6.49 &   86.90$\pm$  4.16 &  21.30$\pm$ 1.98 &	    --       \\
 12    & 16181727-3700597  &   601.00$\pm$ 30.10 &  345.00$\pm$ 19.30 &  291.00$\pm$ 14.30 &  176.00$\pm$  8.45 &  47.20$\pm$ 4.38 &	    --       \\
 13    & 16183414-3715488  &   344.00$\pm$ 17.50 &  214.00$\pm$ 11.00 &  193.00$\pm$  9.41 &  129.00$\pm$  6.09 &  33.60$\pm$ 3.13 &	    --       \\
 14    & 16191403-3747281  & 	99.70$\pm$  5.00 &   85.60$\pm$  4.20 &   67.40$\pm$  3.29 &   59.20$\pm$  2.80 &  73.30$\pm$ 6.78 & 241.00$\pm$28.30\\
 15    & 16192683-3651238  & 	 7.95$\pm$  0.38 &    6.98$\pm$  0.34 &    5.89$\pm$  0.28 &	5.24$\pm$  0.25 &   3.33$\pm$ 0.37 &	    --       \\
 16    & 16194163-3746568  & 	85.60$\pm$  4.32 &   56.40$\pm$  2.73 &   47.70$\pm$  2.27 &   32.60$\pm$  1.56 &  10.20$\pm$ 0.96 &	    --       \\
 17    & 16203160-3730407  &  2070.00$\pm$122.00 & 1240.00$\pm$ 81.50 & 1270.00$\pm$ 73.70 &  910.00$\pm$ 56.50 & 211.00$\pm$19.70 &	    --       \\
 18    & 16205438-3742415  &   614.00$\pm$ 31.90 &  399.00$\pm$ 22.20 &  324.00$\pm$ 16.10 &  200.00$\pm$  9.45 &  46.80$\pm$ 4.34 &	    --       \\
 19    & 16205449-3654430  &   224.00$\pm$ 11.20 &  116.00$\pm$  5.69 &   99.70$\pm$  4.71 &   64.30$\pm$  3.03 &  25.10$\pm$ 2.33 &	    --       \\
 20    & 16212600-3731015  &   757.00$\pm$ 40.90 &  529.00$\pm$ 27.20 &  518.00$\pm$ 26.00 &  387.00$\pm$ 18.20 & 161.00$\pm$14.90 &	    --       \\
 21    & 16212991-3702213  &   189.00$\pm$  9.34 &   99.40$\pm$  4.80 &   83.70$\pm$  3.92 &   57.90$\pm$  2.73 &  56.90$\pm$ 5.27 &	    --       \\
 22    & 16215125-3658489  &  2190.00$\pm$147.00 & 2100.00$\pm$126.00 & 2040.00$\pm$ 98.90 & 2020.00$\pm$111.00 & 976.00$\pm$91.40 &  83.10$\pm$16.20\\
 23    & 16215624-3713167  &   243.00$\pm$ 12.40 &  131.00$\pm$  6.58 &  115.00$\pm$  5.47 &   73.80$\pm$  3.57 &  28.40$\pm$ 2.64 &	    --       \\
 24    & 16215748-3656219  & 	81.90$\pm$  4.03 &   51.40$\pm$  2.44 &   38.30$\pm$  1.81 &   25.90$\pm$  1.21 &   5.99$\pm$ 0.59 &	    --       \\
 25    & 16220163-3716369  & 	65.00$\pm$  3.16 &   42.30$\pm$  2.08 &   35.30$\pm$  1.67 &   24.80$\pm$  1.20 &   8.24$\pm$ 0.79 &	    --       \\
 26    & 16221789-3658165  &   342.00$\pm$ 17.30 &  195.00$\pm$  9.72 &  168.00$\pm$  8.19 &  108.00$\pm$  5.13 &  34.40$\pm$ 3.20 &	    --       \\
 27    & 16223532-3706530  & 	83.70$\pm$  4.10 &   47.30$\pm$  2.29 &   38.70$\pm$  1.82 &   24.00$\pm$  1.13 &   5.59$\pm$ 0.55 &	    --       \\
 28    & 16224265-3721232  &   361.00$\pm$ 18.60 &  274.00$\pm$ 13.90 &  247.00$\pm$ 12.50 &  261.00$\pm$ 13.00 & 198.00$\pm$18.30 &	    --       \\
 29    & 16225309-3724374  &   298.00$\pm$ 14.80 &  234.00$\pm$ 11.70 &  202.00$\pm$  9.70 &  135.00$\pm$  6.43 &  22.40$\pm$ 2.09 &  81.00$\pm$17.60\\
 30    & 16230368-3729411  & 	66.40$\pm$  3.19 &   41.70$\pm$  2.05 &   35.50$\pm$  1.67 &   31.20$\pm$  1.51 &  18.60$\pm$ 1.73 &	    --       \\
 31    & 16232050-3741193  &   183.00$\pm$  9.33 &  165.00$\pm$  7.97 &  139.00$\pm$  6.64 &  125.00$\pm$  6.00 &  97.60$\pm$ 9.08 &	    --       \\
 32    & 16232638-3752082  &   324.00$\pm$ 16.50 &  247.00$\pm$ 12.80 &  233.00$\pm$ 11.80 &  174.00$\pm$  8.28 &  49.60$\pm$ 4.61 &	    --       \\
 33    & 16233246-3752361  &   568.00$\pm$ 28.80 &  323.00$\pm$ 16.60 &  285.00$\pm$ 13.60 &  192.00$\pm$  9.13 &  70.90$\pm$ 6.58 &	    --       \\
 34    & 16233857-3735169  &   149.00$\pm$  7.42 &   95.90$\pm$  4.79 &   74.60$\pm$  3.53 &   50.00$\pm$  2.38 &  14.90$\pm$ 1.39 &	    --       \\
 35    & 16233865-3731388  &   143.00$\pm$  6.89 &   75.90$\pm$  3.74 &   70.50$\pm$  3.32 &   51.60$\pm$  2.44 &  31.40$\pm$ 2.92 &	    --       \\
 36    & 16234903-3727291  & 	 2.10$\pm$  0.10 &    2.67$\pm$  0.13 &    3.07$\pm$  0.15 &	3.36$\pm$  0.17 &   6.46$\pm$ 0.67 &	    --       \\
 37    & 16235052-3756560  &  2420.00$\pm$175.00 & 2130.00$\pm$ 98.60 & 1790.00$\pm$ 88.80 & 1240.00$\pm$ 63.20 & 284.00$\pm$26.30 &	    --       \\
 38    & 16235907-3753528  &  2510.00$\pm$143.00 & 1610.00$\pm$ 80.20 & 1410.00$\pm$ 68.30 &  950.00$\pm$ 45.00 & 319.00$\pm$30.60 &	    --       \\
 39    & 16242666-3757407  &   410.00$\pm$ 20.80 &  214.00$\pm$ 11.00 &  202.00$\pm$  9.78 &  129.00$\pm$  6.21 &  56.20$\pm$ 5.21 &	    --       \\
 40    & 16243230-3754398  &   855.00$\pm$ 43.30 &  607.00$\pm$ 30.00 &  533.00$\pm$ 26.70 &  428.00$\pm$ 20.30 & 170.00$\pm$15.70 &	    --       \\
 41    & 16250690-3803214  &  1640.00$\pm$ 88.20 &  941.00$\pm$ 50.50 &  820.00$\pm$ 39.90 &  539.00$\pm$ 25.90 & 178.00$\pm$16.50 &	    --       \\
 42    & 16182188-3730299  & 	11.20$\pm$  0.91 &    5.57$\pm$  0.43 &    4.09$\pm$  0.28 &	3.05$\pm$  0.19 &   1.73$\pm$ 0.31 &  73.30$\pm$10.90\\
 43    & 16182852-3739386  & 	 7.74$\pm$  0.56 &    4.28$\pm$  0.26 &    3.05$\pm$  0.19 &	2.00$\pm$  0.11 &   1.18$\pm$ 0.24 &	    --       \\
\tableline
\end{tabular}
\end{small}
\end{center}
\end{table*}

\begin{table*}
\begin{center}
\begin{footnotesize}
\caption{IRAC and MIPS observed fluxes for the YSO candidates in Lupus VI. \label{flux_LupVI}}
\begin{tabular}{llllllll}
\tableline
\tableline
  ID & Name/Position   &  3.6 $\mu$m     &   4.5 $\mu$m       &   5.8 $\mu$m        &   8.0 $\mu$m  &  24 $\mu$m   &     70 $\mu$m   \\
     & (SSTgbsJ)       & (mJy)               &  (mJy)         &    (mJy)            &    (mJy)      &    (mJy)     &     (mJy)       \\
\hline
  1    & 16200205-4137264   &  727.00$\pm$ 37.20 &  428.00$\pm$ 22.70 &   360.00$\pm$ 17.20  &  231.00$\pm$ 10.90 &   90.00$\pm$  8.38 &	 --	 \\
  2    & 16200950-4126003   & 1400.00$\pm$ 71.90 &  877.00$\pm$ 47.00 &   788.00$\pm$ 40.20  &  520.00$\pm$ 25.00 &  161.00$\pm$ 14.90 &	 --	 \\
  3    & 16210869-4116025   &	80.50$\pm$  3.96 &   50.40$\pm$  2.42 &    40.00$\pm$  1.88  &   28.10$\pm$  1.32 &    6.71$\pm$  0.65 &	 --	 \\
  4    & 16211125-4120172   &	47.60$\pm$  2.50 &   28.90$\pm$  1.38 &    23.80$\pm$  1.14  &   16.20$\pm$  0.77 &    6.17$\pm$  0.62 &	 --	 \\
  5    & 16213597-4121525   &  825.00$\pm$ 43.40 &  455.00$\pm$ 23.30 &   381.00$\pm$ 18.60  &  234.00$\pm$ 12.30 &   60.30$\pm$  5.58 &	 --	 \\
  6    & 16213962-4109135   &  141.00$\pm$  7.11 &   96.30$\pm$  4.76 &    76.00$\pm$  3.59  &   52.10$\pm$  2.45 &   12.80$\pm$  1.20 &	 --	 \\
  7    & 16214077-4122218   &  158.00$\pm$  7.92 &   86.30$\pm$  4.38 &    73.60$\pm$  3.45  &   48.60$\pm$  2.31 &   19.20$\pm$  1.79 &	 --	 \\
  8    & 16214973-4107017   &  192.00$\pm$  9.39 &  134.00$\pm$  6.60 &   108.00$\pm$  5.09  &   74.80$\pm$  3.59 &   24.00$\pm$  2.23 &	 --	 \\
  9    & 16215797-4131099   &  167.00$\pm$  8.46 &  103.00$\pm$  5.23 &    86.40$\pm$  4.09  &   55.70$\pm$  2.70 &   12.20$\pm$  1.14 &	 --	 \\
 10    & 16222724-4100410   &  198.00$\pm$  9.83 &  114.00$\pm$  5.62 &    91.30$\pm$  4.28  &   60.20$\pm$  2.84 &   18.60$\pm$  1.73 &	 --	 \\
 11    & 16222844-4113424   &  252.00$\pm$ 12.90 &  149.00$\pm$  7.15 &   125.00$\pm$  5.95  &   80.60$\pm$  3.82 &   24.60$\pm$  2.28 &	 --	 \\
 12    & 16222966-4123457   & 1840.00$\pm$206.00 & 2670.00$\pm$168.00 &  3480.00$\pm$220.00  & 2820.00$\pm$136.00 & 1060.00$\pm$101.00 &	 --	 \\
 13    & 16223341-4103179   &  822.00$\pm$ 45.10 &  495.00$\pm$ 25.00 &   426.00$\pm$ 20.40  &  287.00$\pm$ 13.60 &  121.00$\pm$ 11.20 &	 --	 \\
 14    & 16230307-4021119   &  651.00$\pm$ 40.60 &  379.00$\pm$ 20.10 &   377.00$\pm$ 20.10  &  231.00$\pm$ 11.10 &   82.70$\pm$  7.68 &	 --	 \\
 15    & 16231101-4117103   &  167.00$\pm$  8.61 &  115.00$\pm$  5.88 &    92.10$\pm$  4.35  &   61.60$\pm$  2.92 &   17.90$\pm$  1.67 &	 --	 \\
 16    & 16231476-4017151   &  220.00$\pm$ 11.70 &  136.00$\pm$  6.71 &   115.00$\pm$  5.56  &   81.50$\pm$  3.89 &   28.60$\pm$  2.67 &	 --	 \\
 17    & 16231598-4104009   & 1800.00$\pm$126.00 & 1310.00$\pm$ 70.60 &  1120.00$\pm$ 55.30  &  736.00$\pm$ 35.40 &  241.00$\pm$ 22.50 &	 --	 \\
 18    & 16231844-4019439   & 1490.00$\pm$147.00 & 2080.00$\pm$107.00 &  1660.00$\pm$ 87.60  & 1170.00$\pm$ 55.30 &  210.00$\pm$ 19.50 &	 --	 \\
 19    & 16232809-4015368   &	37.70$\pm$  1.87 &   22.80$\pm$  1.09 &    16.70$\pm$  0.87  &    9.83$\pm$  0.47 &   31.10$\pm$  2.90 & 771.00$\pm$81.10\\
 20    & 16233735-4014490   &  136.00$\pm$  6.91 &   70.80$\pm$  3.64 &    60.50$\pm$  2.93  &   38.50$\pm$  1.82 &   11.40$\pm$  1.07 &	 --	 \\
 21    & 16234486-4107493   &  595.00$\pm$ 31.30 &  470.00$\pm$ 27.30 &   407.00$\pm$ 20.20  &  302.00$\pm$ 15.40 &   98.80$\pm$  9.16 &	 --	 \\
 22    & 16234903-4026176   &  164.00$\pm$  8.15 &  113.00$\pm$  5.66 &    97.20$\pm$  4.59  &   77.90$\pm$  3.70 &   25.20$\pm$  2.34 &	 --	 \\
 23    & 16235138-4010324   &  262.00$\pm$ 12.80 &  157.00$\pm$  7.80 &   136.00$\pm$  6.44  &   90.40$\pm$  4.26 &   20.20$\pm$  1.87 &	 --	 \\
 24    & 16242396-3952100   & 1020.00$\pm$ 51.30 &  552.00$\pm$ 31.80 &   527.00$\pm$ 25.00  &  332.00$\pm$ 15.90 &   91.80$\pm$  8.56 &  45.10$\pm$ 8.36\\
 25    & 16242615-4011026   &  232.00$\pm$ 11.90 &  148.00$\pm$  7.77 &   123.00$\pm$  5.85  &   81.80$\pm$  3.96 &   20.00$\pm$  1.85 &	 --	 \\
 26    & 16244169-4004196   &	99.10$\pm$  5.10 &   59.40$\pm$  2.92 &    45.70$\pm$  2.20  &   29.90$\pm$  1.41 &    8.05$\pm$  0.76 &	 --	 \\
 27    & 16244645-3956150   &	56.10$\pm$  2.72 &   51.80$\pm$  2.60 &    47.60$\pm$  2.23  &   37.40$\pm$  1.79 &   45.00$\pm$  4.16 &	 --	 \\
 28    & 16245178-3956326   &	42.90$\pm$  2.13 &   60.30$\pm$  3.07 &    86.00$\pm$  4.06  &  104.00$\pm$  5.18 &  211.00$\pm$ 19.50 & 110.00$\pm$15.60\\
 29    & 16245564-3949147   &  109.00$\pm$  6.76 &   99.40$\pm$  4.81 &    73.00$\pm$  3.59  &   52.30$\pm$  2.46 &   12.80$\pm$  1.20 &	 --	 \\
 30    & 16245590-4011282   & 1430.00$\pm$101.00 &  949.00$\pm$ 54.00 &   954.00$\pm$ 50.50  &  674.00$\pm$ 33.50 &  178.00$\pm$ 16.50 &	 --	 \\
 31    & 16245681-4008238   &  250.00$\pm$ 15.90 &  214.00$\pm$ 10.90 &   184.00$\pm$  8.90  &  119.00$\pm$  5.93 &   24.40$\pm$  2.27 &	 --	 \\
 32    & 16255246-4018484   &  123.00$\pm$  9.05 &   79.60$\pm$  4.95 &    73.80$\pm$  3.74  &   46.80$\pm$  2.37 &   12.80$\pm$  1.20 &	 --	 \\
 33    & 16255837-4009581   &  777.00$\pm$ 42.80 &  423.00$\pm$ 26.50 &   406.00$\pm$ 20.50  &  263.00$\pm$ 13.20 &  105.00$\pm$  9.73 &	 --	 \\
 34    & 16260853-4017441   &  459.00$\pm$ 30.80 &  241.00$\pm$ 14.90 &   246.00$\pm$ 14.30  &  153.00$\pm$  7.66 &   33.20$\pm$  3.10 &	 --	 \\
 35    & 16261339-3949542   &  215.00$\pm$ 12.70 &  160.00$\pm$  8.55 &   134.00$\pm$  6.98  &  101.00$\pm$  5.01 &   37.00$\pm$  3.44 &	 --	 \\
 36    & 16261365-4012186   &  636.00$\pm$ 36.00 &  286.00$\pm$ 21.70 &   317.00$\pm$ 16.70  &  214.00$\pm$ 11.20 &   52.50$\pm$  4.87 &	 --	 \\
 37    & 16262551-3944472   &	47.00$\pm$  2.64 &   32.00$\pm$  1.65 &    25.20$\pm$  1.29  &   29.40$\pm$  1.43 &   15.30$\pm$  1.42 &	 --	 \\
 38    & 16263114-3946153   &	 2.76$\pm$  0.16 &    1.70$\pm$  0.09 &     1.31$\pm$  0.08  &    0.85$\pm$  0.05 &    0.84$\pm$  0.22 &	 --	 \\
 39    & 16263155-4004557   &  597.00$\pm$ 32.80 &  375.00$\pm$ 21.70 &   313.00$\pm$ 17.30  &  209.00$\pm$ 10.30 &   50.60$\pm$  4.69 &	 --	 \\
 40    & 16263925-4015272   &  393.00$\pm$ 24.80 &  263.00$\pm$ 16.10 &   310.00$\pm$ 16.30  &  212.00$\pm$ 10.60 &  115.00$\pm$ 10.60 &	 --	 \\
 41    & 16265364-3954594   &  283.00$\pm$ 17.30 &  223.00$\pm$ 12.50 &   200.00$\pm$ 10.50  &  136.00$\pm$  6.73 &   21.00$\pm$  1.95 &	 --	 \\
 42    & 16270309-3944139   &  293.00$\pm$ 20.00 &  255.00$\pm$ 12.90 &   199.00$\pm$ 11.70  &  154.00$\pm$  7.27 &   32.90$\pm$  3.06 &	 --	 \\
 43    & 16273175-3956013   &	70.70$\pm$  3.66 &   41.20$\pm$  2.15 &    35.50$\pm$  1.68  &   22.40$\pm$  1.10 &    5.59$\pm$  0.56 &	 --	 \\
 44    & 16275054-3948100   &  471.00$\pm$ 27.70 &  281.00$\pm$ 13.80 &   239.00$\pm$ 11.40  &  159.00$\pm$  7.51 &   57.40$\pm$  5.31 &	 --	 \\
 45    & 16282647-3925428   &  126.00$\pm$  6.61 &  139.00$\pm$  6.88 &   188.00$\pm$  9.34  &  272.00$\pm$ 13.00 &  175.00$\pm$ 16.30 &	 --	 \\
\tableline
\end{tabular}
\end{footnotesize}
\end{center}
\end{table*}

\begin{table*}
\begin{center}
\begin{footnotesize}
\caption{NOMAD, DENIS and 2MASS observed magnitudes for the YSO candidates in Lupus V. \label{tab_mag_LupV}}
\begin{tabular}{lllllllll}
\tableline
\tableline
  ID & Name/Position   &     B$^\dag$       &    V$^\dag$   &     R$^\dag$      &  I &  J    &  H   & K$_S$  \\  
     & (SSTgbsJ)       &                    &               &                   &    &       &      &         \\                    
\hline
  1    & 16163197-3704563  &  17.85   &  15.31  &   --     &   --      		   &   7.06$\pm$0.02 &  5.72$\pm$0.03 &  5.10$\pm$0.02 \\
  2    & 16164035-3725054  &  18.49   &  --     &   --     &   --      		   &   8.46$\pm$0.02 &  7.28$\pm$0.04 &  6.43$\pm$0.03 \\
  3    & 16164198-3650456  &  19.30   &  --     &  17.47   &   --      		   &  14.41$\pm$0.03 & 13.58$\pm$0.03 & 13.29$\pm$0.03 \\
  4    & 16171811-3646307  &  19.89   &  --     &  18.10   &   16.05$\pm$0.06	   &  13.66$\pm$0.02 & 12.99$\pm$0.02 & 12.66$\pm$0.03 \\
  5    & 16172474-3657333  &  17.25   &  16.44  &   --     &   13.24$\pm$0.03	   &  10.56$\pm$0.04 &  9.61$\pm$0.04 &  9.09$\pm$0.03 \\
  6    & 16172485-3657407  &  16.96   &  15.62  &  14.29   &   13.02$\pm$0.03	   &  11.47$\pm$0.03 & 10.83$\pm$0.03 & 10.59$\pm$0.02 \\
  7    & 16175383-3645207  &  17.61   &  --     &   9.85   &    8.59$\pm$0.04	   &   5.59$\pm$0.02 &  4.48$\pm$0.23 &  4.11$\pm$0.23 \\
  8    & 16175958-3719083  &  --      &  --     &  18.05   &   14.43$\pm$0.05	   &  10.10$\pm$0.02 &  8.85$\pm$0.02 &  8.18$\pm$0.03 \\
  9    & 16180425-3710277  &  18.38   &  16.82  &  14.86   &   14.18$\pm$0.04	   &  12.05$\pm$0.03 & 10.92$\pm$0.04 & 10.50$\pm$0.02 \\
 10    & 16180702-3706544  &  --      &  17.74  &  --	   &   11.39$\pm$0.02	   &   7.34$\pm$0.03 &  6.04$\pm$0.03 &  5.33$\pm$0.02 \\
 11    & 16180915-3725364  &  --      &  17.32  &  15.91   &   12.71$\pm$0.02	   &   9.69$\pm$0.02 &  8.40$\pm$0.05 &  7.94$\pm$0.03 \\
 12    & 16181727-3700597  &  18.93   &  17.18  &  15.68   &   12.10$\pm$0.02	   &   8.89$\pm$0.02 &  7.61$\pm$0.03 &  7.00$\pm$0.02 \\
 13    & 16183414-3715488  &  19.44   &  17.74  &  --	   &   13.62$\pm$0.04	   &   9.64$\pm$0.02 &  8.30$\pm$0.06 &  7.72$\pm$0.05 \\
 14    & 16191403-3747281  &  15.72   &  14.44  &  13.36   &   13.13$\pm$0.03	   &  11.69$\pm$0.02 & 10.63$\pm$0.02 &  9.92$\pm$0.02 \\
 15    & 16192683-3651238  &  17.95   &  --     &  16.93   &   15.41$\pm$0.05	   &  13.13$\pm$0.03 & 12.56$\pm$0.03 & 12.16$\pm$0.03 \\
 16    & 16194163-3746568  &  19.45   &  --     &  --	   &   14.27$\pm$0.03	   &  10.98$\pm$0.02 &  9.77$\pm$0.02 &  9.25$\pm$0.02 \\
 17    & 16203160-3730407  &  --      &  17.32  &  --	   &   12.03$\pm$1.00	   &   7.64$\pm$0.02 &  6.42$\pm$0.02 &  5.74$\pm$0.02 \\
 18    & 16205438-3742415  &  18.27   &  --     &  15.23   &   12.16$\pm$1.00	   &   8.72$\pm$0.02 &  7.53$\pm$0.03 &  6.98$\pm$0.02 \\
 19    & 16205449-3654430  &  18.47   &  --     &  --	   &   13.11$\pm$1.00	   &   9.82$\pm$0.02 &  8.56$\pm$0.02 &  8.02$\pm$0.02 \\
 20    & 16212600-3731015  &  --      &  --     &  16.69   &   -- 		   &   9.98$\pm$0.02 &  8.47$\pm$0.02 &  7.54$\pm$0.06 \\
 21    & 16212991-3702213  &  18.38   &  16.07  &  14.92   &   12.61$\pm$0.03	   &   9.94$\pm$0.02 &  8.78$\pm$0.02 &  8.26$\pm$0.02 \\
 22    & 16215125-3658489  &  18.35   &  16.50  &  --	   &   11.52$\pm$0.03	   &   7.08$\pm$0.02 &  5.90$\pm$0.03 &  5.19$\pm$0.02 \\
 23    & 16215624-3713167  &  19.19   &  --     &  --	   &   13.64$\pm$0.03	   &   9.78$\pm$0.02 &  8.55$\pm$0.02 &  7.97$\pm$0.02 \\
 24    & 16215748-3656219  &  18.38   &  16.37  &  14.95   &   13.24$\pm$0.03	   &  10.75$\pm$0.02 &  9.55$\pm$0.02 &  9.13$\pm$0.02 \\
 25    & 16220163-3716369  &  18.42   &  17.28  &  16.20   &   13.61$\pm$0.03	   &  11.16$\pm$0.02 &  9.96$\pm$0.02 &  9.46$\pm$0.02 \\
 26    & 16221789-3658165  &  18.73   &  --     &  --	   &   13.07$\pm$0.03	   &   9.43$\pm$0.02 &  8.23$\pm$0.02 &  7.64$\pm$0.02 \\
 27    & 16223532-3706530  &  19.33   &  --     &  17.17   &   14.18$\pm$0.03	   &  10.90$\pm$0.02 &  9.63$\pm$0.02 &  9.10$\pm$0.02 \\
 28    & 16224265-3721232  &  18.82   &  --     &  16.08   &   13.21$\pm$0.03	   &   9.77$\pm$0.02 &  8.43$\pm$0.02 &  7.82$\pm$0.03 \\
 29    & 16225309-3724374  &  18.33   &  16.44  &  --	   &	--	           &   9.96$\pm$0.02 &  8.73$\pm$0.02 &  8.13$\pm$0.03 \\
 30    & 16230368-3729411  &  17.88   &  17.41  &  16.02   &   13.98$\pm$0.03	   &  11.05$\pm$0.02 &  9.84$\pm$0.02 &  9.36$\pm$0.02 \\
 31    & 16232050-3741193  &  --      &  --     &  18.09   &   14.09$\pm$0.05	   &  10.84$\pm$0.02 &  9.67$\pm$0.02 &  8.9 $\pm$0.02 \\
 32    & 16232638-3752082  &  --      &  --     &  17.31   &	--  	           &  10.31$\pm$0.02 &  9.10$\pm$0.02 &  8.37$\pm$0.02 \\
 33    & 16233246-3752361  &  18.35   &  16.67  &  --	   &   11.88$\pm$0.04	   &   8.82$\pm$0.02 &  7.55$\pm$0.02 &  7.00$\pm$0.02 \\
 34    & 16233857-3735169  &  18.10   &  17.15  &  --	   &   12.84$\pm$0.05	   &  10.13$\pm$0.02 &  8.96$\pm$0.02 &  8.50$\pm$0.02 \\
 35    & 16233865-3731388  &  18.58   &  17.74  &  15.80   &   13.10$\pm$0.05	   &  10.25$\pm$0.02 &  9.08$\pm$0.02 &  8.55$\pm$0.02 \\
 36    & 16234903-3727291  &  --      &  --     &  --	   &   --  		   &  --  	     &  --            & 15.08$\pm$0.16 \\
 37    & 16235052-3756560  &  18.31   &  16.11  &  14.41   &   10.94$\pm$0.05	   &   6.97$\pm$0.02 &  5.66$\pm$0.03 &  5.08$\pm$0.01 \\
 38    & 16235907-3753528  &  18.38   &  15.99  &  14.28   &   10.74$\pm$0.06	   &   7.14$\pm$0.02 &  5.87$\pm$0.05 &  5.28$\pm$0.03 \\
 39    & 16242666-3757407  &  19.57   &  --     &  --	   &   13.52$\pm$0.03	   &   9.27$\pm$0.02 &  7.97$\pm$0.05 &  7.35$\pm$0.02 \\
 40    & 16243230-3754398  &  17.49   &  --     &  --	   &   11.76$\pm$0.02	   &   8.43$\pm$0.03 &  7.20$\pm$0.03 &  6.70$\pm$0.02 \\
 41    & 16250690-3803214  &  18.38   &  16.69  &  14.99   &   11.49$\pm$0.02	   &   7.69$\pm$0.02 &  6.44$\pm$0.02 &  5.87$\pm$0.02 \\
 42    & 16182188-3730299  &  17.86   &  --     &  --	   &   14.54$\pm$0.05	   &  12.79$\pm$0.10 & 11.73$\pm$0.11 & 11.31$\pm$0.09 \\
 43    & 16182852-3739386  &  17.40   &  --     &  --	   &   15.36$\pm$0.07	   &  13.32$\pm$0.07 & 12.30$\pm$0.09 & 11.77$\pm$0.06 \\
\tableline
\end{tabular}
\end{footnotesize}
\end{center}
$^\dag$ Photometric uncertainties of the NOMAD magnitudes are of the order of 1-2\% \citep{Zacharias2005}.\\
\end{table*}

\begin{table*}
\begin{center}
\begin{footnotesize}
\caption{NOMAD, DENIS and 2MASS observed magnitudes for the YSO candidates in Lupus VI. \label{tab_mag_LupVI}}
\begin{tabular}{lllllllll}
\tableline
\tableline
  ID & Name/Position   &     B$^\dag$       &    V$^\dag$   &     R$^\dag$      &  I &  J    &  H   & K$_S$  \\  
     & (SSTgbsJ)       &             &        &            &    &       &     &         \\                    
\hline
  1    & 16200205-4137264    &  17.31	&    --     &  --	& 12.06$\pm$0.02  &  8.53$\pm$0.03  &  7.29$\pm$0.03  &  6.73$\pm$0.03 \\ 
  2    & 16200950-4126003    &  17.56	&    --     &  --	& 12.46$\pm$0.02  &  8.29$\pm$0.02  &  6.91$\pm$0.03  &  6.28$\pm$0.01 \\ 
  3    & 16210869-4116025    &  17.33	&    --     &  16.29	& 13.72$\pm$1.00  & 10.83$\pm$0.02  &  9.62$\pm$0.02  &  9.14$\pm$0.02 \\ 
  4    & 16211125-4120172    &  --	&    --     &  18.71	& 15.40$\pm$1.00  & 11.80$\pm$0.02  & 10.48$\pm$0.02  &  9.94$\pm$0.02 \\ 
  5    & 16213597-4121525    &  17.74	&    16.26  &  14.94	& 11.86$\pm$0.03  &  8.35$\pm$0.02  &  7.12$\pm$0.02  &  6.55$\pm$0.02 \\ 
  6    & 16213962-4109135    &  18.84	&    --     &  16.17	& 13.31$\pm$0.03  & 10.29$\pm$0.02  &  9.05$\pm$0.02  &  8.58$\pm$0.02 \\ 
  7    & 16214077-4122218    &  17.89	&    --     &  15.98	& 13.67$\pm$0.03  & 10.26$\pm$0.02  &  9.03$\pm$0.02  &  8.51$\pm$0.02 \\ 
  8    & 16214973-4107017    &  18.03	&    --     &  15.35	& 13.05$\pm$0.03  & 10.31$\pm$0.02  &  9.12$\pm$0.02  &  8.50$\pm$0.02 \\ 
  9    & 16215797-4131099    &  17.27	&    16.98  &  15.22	& 12.77$\pm$0.03  & 10.08$\pm$0.02  &  8.80$\pm$0.02  &  8.32$\pm$0.02 \\ 
 10    & 16222724-4100410    &  18.00	&    --     &  16.41	& 13.48$\pm$0.03  & 10.03$\pm$0.02  &  8.80$\pm$0.02  &  8.24$\pm$0.02 \\ 
 11    & 16222844-4113424    &  --	&    --     &  16.74	& 14.10$\pm$0.03  &  9.98$\pm$0.02  &  8.54$\pm$0.03  &  7.91$\pm$0.02 \\ 
 12    & 16222966-4123457    &  --	&    --     &  --	& 11.81$\pm$0.02  &  7.18$\pm$0.02  &  5.92$\pm$0.03  &  5.07$\pm$0.02 \\ 
 13    & 16223341-4103179    &  18.25	&    --     &  --	& 13.19$\pm$0.03  &  8.61$\pm$0.03  &  7.22$\pm$0.02  &  6.62$\pm$0.03 \\ 
 14    & 16230307-4021119    &  17.93	&    --     &  --	& 13.06$\pm$0.02  &  9.32$\pm$0.02  &  8.04$\pm$0.03  &  7.38$\pm$0.03 \\ 
 15    & 16231101-4117103    &  18.45	&    --     &  17.19	& 13.86$\pm$0.03  & 10.36$\pm$0.03  &  8.97$\pm$0.02  &  8.37$\pm$0.02 \\ 
 16    & 16231476-4017151    &  18.47	&    --     &  16.96	& 13.71$\pm$0.03  &  9.94$\pm$0.02  &  8.65$\pm$0.02  &  8.03$\pm$0.02 \\ 
 17    & 16231598-4104009    &  16.00	&    16.14  &  13.11	& 11.46$\pm$0.03  &  7.27$\pm$0.02  &  5.98$\pm$0.02  &  5.34$\pm$0.02 \\ 
 18    & 16231844-4019439    &  15.48	&    13.40  &  12.35	& 10.56$\pm$0.06  &  6.98$\pm$0.02  &  5.76$\pm$0.03  &  5.25$\pm$0.02 \\ 
 19    & 16232809-4015368    &  --	&    17.22  &  --	& 14.81$\pm$0.06  & 11.82$\pm$0.03  & 10.50$\pm$0.03  & 10.05$\pm$0.03 \\ 
 20    & 16233735-4014490    &  --	&    --     &  18.69	& 15.15$\pm$0.06  & 10.75$\pm$0.03  &  9.38$\pm$0.02  &  8.73$\pm$0.02 \\ 
 21    & 16234486-4107493    &  --	&    --     &  --	& 12.86$\pm$0.05  &  9.83$\pm$0.02  &  8.55$\pm$0.04  &  7.68$\pm$0.03 \\ 
 22    & 16234903-4026176    &  18.07	&    --     &  17.12	& 14.15$\pm$0.05  & 10.55$\pm$0.02  &  9.23$\pm$0.02  &  8.64$\pm$0.02 \\ 
 23    & 16235138-4010324    &  18.07	&    --     &  15.96	& 13.77$\pm$0.05  &  9.84$\pm$0.02  &  8.63$\pm$0.05  &  8.05$\pm$0.04 \\ 
 24    & 16242396-3952100    &  17.15	&    --     &  --	& 11.99$\pm$0.02  &  8.41$\pm$0.02  &  7.15$\pm$0.03  &  6.56$\pm$0.02 \\ 
 25    & 16242615-4011026    &  --	&    17.20  &  --	& 13.44$\pm$0.03  & 10.16$\pm$0.02  &  8.91$\pm$0.02  &  8.29$\pm$0.04 \\ 
 26    & 16244169-4004196    &  17.88	&    17.74  &  16.81	& 13.85$\pm$0.03  & 10.84$\pm$0.02  &  9.52$\pm$0.02  &  8.97$\pm$0.02 \\ 
 27    & 16244645-3956150    &  --	&    --     &  --	& --   		  & 13.90$\pm$0.03  & 11.84$\pm$0.02  & 10.65$\pm$0.02 \\ 
 28    & 16245178-3956326    &  --	&    --     &  --	& --    	  & 16.20$\pm$0.12  & 14.24$\pm$0.06  & 12.69$\pm$0.03 \\ 
 29    & 16245564-3949147    &  17.27	&    16.91  &  15.93	& 13.08$\pm$0.03  & 10.42$\pm$0.03  &  9.30$\pm$0.02  &  8.72$\pm$0.02 \\ 
 30    & 16245590-4011282    &  --	&    --     &  15.15	& 12.07$\pm$0.02  &  7.95$\pm$0.02  &  6.72$\pm$0.06  &  6.00$\pm$0.02 \\ 
 31    & 16245681-4008238    &  17.39	&    --     &  15.31	& 12.40$\pm$0.02  &  9.51$\pm$0.02  &  8.26$\pm$0.05  &  7.71$\pm$0.03 \\ 
 32    & 16255246-4018484    &  18.14	&    --     &  17.39	& 13.75$\pm$0.03  & 10.28$\pm$0.02  &  8.98$\pm$0.02  &  8.39$\pm$0.03 \\ 
 33    & 16255837-4009581    &  18.53	&    --     &  --	& 13.19$\pm$0.02  &  8.72$\pm$0.02  &  7.33$\pm$0.05  &  6.66$\pm$0.03 \\ 
 34    & 16260853-4017441    &  16.86	&    16.68  &  15.28	& 12.12$\pm$0.02  &  8.97$\pm$0.02  &  7.67$\pm$0.05  &  7.08$\pm$0.02 \\ 
 35    & 16261339-3949542    &  17.74	&    --     &  14.83	& 13.30$\pm$0.02  &  9.94$\pm$0.02  &  8.83$\pm$0.04  &  8.18$\pm$0.04 \\ 
 36    & 16261365-4012186    &  17.95	&    17.11  &  15.81	& 12.57$\pm$0.02  &  8.89$\pm$0.03  &  7.46$\pm$0.05  &  6.80$\pm$0.02 \\ 
 37    & 16262551-3944472    &  17.15	&    --     &  15.44	& 13.71$\pm$0.03  & 11.31$\pm$0.02  & 10.10$\pm$0.02  &  9.66$\pm$0.02 \\ 
 38    & 16263114-3946153    &  --	&    --     &  --	& 15.43$\pm$0.05  & 13.78$\pm$0.06  & --              & --             \\ 
 39    & 16263155-4004557    &  18.00	&    --     &  15.84	& 12.63$\pm$0.03  &  9.05$\pm$0.03  &  7.65$\pm$0.04  &  7.03$\pm$0.02 \\ 
 40    & 16263925-4015272    &  --	&    --     &  --	& 17.10$\pm$0.12  & 11.13$\pm$0.02  &  9.42$\pm$0.02  &  8.41$\pm$0.02 \\ 
 41    & 16265364-3954594    &  18.08	&    16.56  &  --	& 12.25$\pm$0.02  &  9.94$\pm$0.02  &  8.72$\pm$0.02  &  8.09$\pm$0.02 \\ 
 42    & 16270309-3944139    &  --	&    --     &  --	& 12.57$\pm$0.03  & 10.01$\pm$0.02  &  8.80$\pm$0.02  &  8.11$\pm$0.02 \\ 
 43    & 16273175-3956013    &  18.35	&    --     &  16.83	& 14.44$\pm$0.03  & 11.3 $\pm$0.02  &  9.99$\pm$0.02  &  9.44$\pm$0.02 \\ 
 44    & 16275054-3948100    &  17.84	&    --     &  16.20	& 12.78$\pm$0.03  &  9.26$\pm$0.02  &  7.97$\pm$0.05  &  7.39$\pm$0.02 \\ 
 45    & 16282647-3925428    &  16.74	&    --     &  14.25	& --              & 11.37$\pm$0.02  & 10.26$\pm$0.02  &  9.80$\pm$0.02 \\ 
\tableline
\end{tabular}
\end{footnotesize}
\end{center}
$^\dag$Photometric uncertainties of the NOMAD magnitudes are of the order of 1-2\% \citep{Zacharias2005}.
\end{table*}

\begin{table*}
\begin{center}
\begin{footnotesize}
\caption{SED slopes and Lada class for the YSO candidates in Lupus V. \label{tab_param_V}}
\begin{tabular}{llllll}
\tableline
\tableline
  ID  & Name/Position    & $\alpha_{K-24}$ &  $\alpha_{3.6-5.8}$ & $\alpha_{8-24}$ & Lada class  \\\hline
   1  & 16163197-3704563  &   -2.06  &  -0.66  & -2.76    &   III     \\ 
   2  & 16164035-3725054  &   -1.62  &  -1.70  & -1.64    &   III     \\ 
   3* & 16164198-3650456  &   -0.55  &  -2.59  &  2.10    &    II     \\ 
   4  & 16171811-3646307  &   -1.03  &  -1.49  & -0.64    &    II     \\ 
   5  & 16172474-3657333  &   -1.15  &  -1.42  & -0.80    &    II     \\ 
   6  & 16172485-3657407  &   -2.43  &  -2.68  & -1.15    &   III     \\ 
   7  & 16175383-3645207  &   -2.47  &  -0.11  & -3.03    &   III     \\ 
   8  & 16175958-3719083  &   -2.02  &  -2.40  & -1.69    &   III     \\ 
   9  & 16180425-3710277  &   -2.35  &  -2.60  & -2.39    &   III     \\ 
  10  & 16180702-3706544  &   -2.08  &  -1.90  & -2.09    &   III     \\ 
  11  & 16180915-3725364  &   -2.27  &  -2.31  & -2.28    &   III     \\ 
  12  & 16181727-3700597  &   -2.34  &  -2.50  & -2.20    &   III     \\ 
  13  & 16183414-3715488  &   -2.16  &  -2.19  & -2.22    &   III     \\ 
  14  & 16191403-3747281  &   -1.21  &  -1.82  &  -0.81   &    II     \\ 
  15  & 16192683-3651238  &   -1.46  &  -1.63  &  -1.41   &    II     \\ 
  16  & 16194163-3746568  &   -2.09  &  -2.21  & -2.06    &   III     \\ 
  17  & 16203160-3730407  &   -2.11  &  -2.00  & -2.33    &   III     \\ 
  18  & 16205438-3742415  &   -2.32  &  -2.33  & -2.32    &   III     \\ 
  19  & 16205449-3654430  &   -2.22  &  -2.67  & -1.86    &   III     \\ 
  20  & 16212600-3731015  &   -1.69  &  -1.78  & -1.80    &   III     \\ 
  21  & 16212991-3702213  &   -1.86  &  -2.68  & -1.02    &   III     \\ 
  22  & 16215125-3658489  &   -1.54  &  -1.15  &  -1.66   &    II     \\ 
  23  & 16215624-3713167  &   -2.18  &  -2.54  & -1.87    &   III     \\ 
  24  & 16215748-3656219  &   -2.36  &  -2.58  & -2.33    &   III     \\ 
  25  & 16220163-3716369  &   -2.09  &  -2.27  & -2.00    &   III     \\ 
  26  & 16221789-3658165  &   -2.22  &  -2.47  & -2.04    &   III     \\ 
  27  & 16223532-3706530  &   -2.41  &  -2.60  & -2.33    &   III     \\ 
  28  & 16224265-3721232  &   -1.35  &  -1.79  &  -1.25   &    II     \\ 
  29  & 16225309-3724374  &   -2.16  &  -1.81  & -2.63    &   III     \\ 
  30  & 16230368-3729411  &   -1.76  &  -2.30  & -1.47    &   III     \\ 
  31  & 16232050-3741193  &   -1.34  &  -1.58  &  -1.23   &    II     \\ 
  32  & 16232638-3752082  &   -1.80  &  -1.68  & -2.14    &   III     \\ 
  33  & 16233246-3752361  &   -2.14  &  -2.42  & -1.91    &   III     \\ 
  34  & 16233857-3735169  &   -2.23  &  -2.44  & -2.10    &   III     \\ 
  35  & 16233865-3731388  &   -1.91  &  -2.46  & -1.45    &   III     \\ 
  36  & 16234903-3727291  &   -0.35  &  -0.21  &  -0.40   &    II     \\ 
  37  & 16235052-3756560  &   -2.17  &  -1.63  & -2.34    &   III     \\ 
  38  & 16235907-3753528  &   -2.12  &  -2.19  & -1.99    &   III     \\ 
  39  & 16242666-3757407  &   -2.12  &  -2.46  & -1.76    &   III     \\ 
  40  & 16243230-3754398  &   -1.85  &  -1.98  & -1.84    &   III     \\ 
  41  & 16250690-3803214  &   -2.19  &  -2.43  & -2.01    &   III     \\ 
  42  & 16182188-3730299  &   -2.22  &  -3.09  & -1.52    &   III     \\ 
  43  & 16182852-3739386  &   -2.32  &  -2.94  & -1.48    &   III     \\ 
\tableline
\end{tabular}
\end{footnotesize}
\end{center}
$^*$Transition object candidate.\\
\end{table*}

\begin{table*}
\begin{center}
\begin{footnotesize}
\caption{SED slopes and Lada class for the YSO candidates in Lupus VI. \label{tab_param_VI}}
\begin{tabular}{llllll}
\tableline
\tableline
  ID   & Name/Position   & $\alpha_{K-24}$ & $\alpha_{3.6-5.8}$ & $\alpha_{8-24}$ & Lada class    \\
  \hline
  1    & 16200205-4137264   &  -2.17 &   -2.46 &  -1.86 & III  \\
  2    & 16200950-4126003   &  -2.09 &   -2.19 &  -2.07 & III  \\
  3    & 16210869-4116025   &  -2.27 &   -2.45 &  -2.30 & III  \\
  4    & 16211125-4120172   &  -2.09 &   -2.44 &  -1.88 & III  \\
  5    & 16213597-4121525   &  -2.38 &   -2.60 &  -2.23 & III  \\
  6    & 16213962-4109135   &  -2.23 &   -2.29 &  -2.28 & III  \\
  7    & 16214077-4122218   &  -2.16 &   -2.58 &  -1.85 & III  \\
  8    & 16214973-4107017   &  -2.06 &   -2.20 &  -2.03 & III  \\
  9    & 16215797-4131099   &  -2.33 &   -2.37 &  -2.38 & III  \\
 10    & 16222724-4100410   &  -2.26 &   -2.61 &  -2.07 & III  \\
 11    & 16222844-4113424   &  -2.24 &   -2.45 &  -2.08 & III  \\
 12    & 16222966-4123457   &  -1.53 &    0.33 &  -1.89 &  II  \\
 13    & 16223341-4103179   &  -2.07 &   -2.36 &  -1.79 & III  \\
 14    & 16230307-4021119   &  -2.00 &   -2.12 &  -1.93 & III  \\
 15    & 16231101-4117103   &  -2.18 &   -2.24 &  -2.12 & III  \\
 16    & 16231476-4017151   &  -2.10 &   -2.34 &  -1.95 & III  \\
 17    & 16231598-4104009   &  -2.17 &   -1.99 &  -2.02 & III  \\
 18    & 16231844-4019439   &  -2.20 &   -0.80 &  -2.56 & III  \\
 19*   & 16232809-4015368   &  -1.59 &   -2.70 &   0.05 &  II  \\
 20    & 16233735-4014490   &  -2.29 &   -2.67 &  -2.11 & III  \\
 21    & 16234486-4107493   &  -1.80 &   -1.79 &  -2.02 & III  \\
 22    & 16234903-4026176   &  -1.92 &   -2.09 &  -2.03 & III  \\
 23    & 16235138-4010324   &  -2.26 &   -2.36 &  -2.36 & III  \\
 24    & 16242396-3952100   &  -2.22 &   -2.36 &  -2.17 & III  \\
 25    & 16242615-4011026   &  -2.21 &   -2.32 &  -2.28 & III  \\
 26    & 16244169-4004196   &  -2.32 &   -2.61 &  -2.19 & III  \\
 27    & 16244645-3956150   &  -1.12 &   -1.34 &  -0.83 &  II  \\
 28    & 16245178-3956326   &   0.22 &    0.46 &  -0.36 &  F   \\
 29    & 16245564-3949147   &  -2.13 &   -1.85 &  -2.28 & III  \\
 30    & 16245590-4011282   &  -2.05 &   -1.83 &  -2.21 & III  \\
 31    & 16245681-4008238   &  -2.21 &   -1.64 &  -2.44 & III  \\
 32    & 16255246-4018484   &  -2.23 &   -2.05 &  -2.18 & III  \\
 33    & 16255837-4009581   &  -2.10 &   -2.33 &  -1.84 & III  \\
 34    & 16260853-4017441   &  -2.34 &   -2.28 &  -2.39 & III  \\
 35    & 16261339-3949542   &  -1.92 &   -1.98 &  -1.91 & III  \\
 36    & 16261365-4012186   &  -2.28 &   -2.42 &  -2.28 & III  \\
 37    & 16262551-3944472   &  -1.64 &   -2.30 &  -1.59 & III  \\
 38    & 16263114-3946153   &  -2.14 &   -2.55 &  -1.01 & III  \\
 39    & 16263155-4004557   &  -2.26 &   -2.34 &  -2.29 & III  \\
 40    & 16263925-4015272   &  -1.49 &   -1.47 &  -1.56 &  II  \\
 41    & 16265364-3954594   &  -2.18 &   -1.72 &  -2.70 & III  \\
 42    & 16270309-3944139   &  -2.01 &   -1.81 &  -2.40 & III  \\
 43    & 16273175-3956013   &  -2.28 &   -2.42 &  -2.26 & III  \\
 44    & 16275054-3948100   &  -2.10 &   -2.40 &  -1.93 & III  \\
 45    & 16282647-3925428   &  -0.53 &   -0.15 &  -1.40 &  II  \\
\tableline
\end{tabular}
\end{footnotesize}
\end{center}
$^*$Transition object candidate.\\
\end{table*}

\begin{table*}
\begin{center}
\caption{Total number of YSO candidates in the Lupus V and VI
clouds organized by Lada class. We also report for comparison the \cite{Merin2008}
values for Lupus I, III and IV and the values for other clouds observed within the Spitzer c2d Survey. 
This last sample includes the statistics for Cha~II \citep{Alcala2008},
Perseus \citep{Rebull2007,Jorgensen2008,Evans2009a}, Serpens \citep{Harvey2007b}, and Ophiuchus \citep{Padgett2008,Evans2009a}. \label{classes}}
\begin{tabular}{c|c|c|c|c|c|c}
\hline
Lada Class & Lupus I$^\dag$  & Lupus III$^\dag$ & Lupus IV$^\dag$ &  Lupus V & Lupus VI & All c2d clouds$^\ddag$ \\
\hline \hline
I             & 2 (15\%)   & 2 (3\%)      & 1 (8\%)    & 0                      & 0                & 165 (16\%)  \\
Flat       & 3 (23\%)   & 6 (9\%)       & 1 (8\%)    & 0                     & 1 (2\%)      & 123 (12\%)  \\
II           & 6 (47\%)    & 41 (59\%)  & 5 (42\%) & 9  (21\%)       & 5 (11\%)     & 612 (60\%)  \\
III          & 2 (15\%)    & 20 (29\%)  & 5 (42\%) & 34 (79\%)       & 39 (87\%)  & 124 (12\%)  \\
Total    & 13               & 69               & 12           & 43                    & 45                 & 1024               \\
\hline
\end{tabular}
\end{center}
$^\dag$ From  \citet{Merin2008}.\\
$^\ddag$ From  \citet{Evans2009a}; see also their Table~5 for values relative to each c2d cloud. \\
\end{table*}

\begin{table*}
\small \caption{Summary of the clustering analysis versus Lada classes for Lupus V and VI (see Figure~\ref{clust_fig}).  \label{clust_tab}}
\begin{tabular}{llllllllllll}\hline\hline
Region & Total YSO & Class I & Flat & Class II & Class III & Cloud mass  & Volume   & $\langle A_V \rangle$ & SFE$^a$  &  SFE$^b$ \\%&   SFE$^c$ \\
              & candidates &              &          &               &                & $[M_\odot]$ & [pc$^3$] &                                         & $[\%]$        & $[\%]$       \\%&  $[\%]$  \\
\hline
Va  &           22 &            0 &            0 &            4 &           18 &        330       &        8.6       &       3.1     &        3.2 &  5.4  \\ %& 7.9 \\
Vb  &           13 &            0 &            0 &            3 &           10 &        210       &        4.9       &       2.9     &        2.9 &  5.0 \\  %& 7.4 \\
VIa &           28 &            0 &            1 &            3 &           24 &        290       &        9.4       &       2.6     &        4.6 &  7.9  \\ % & 11.0 \\
VIb &           15 &            0 &            0 &            1 &           14 &        160       &        3.5       &       2.8     &        4.4 &  7.1 \\  %& 10.9\\
\\\hline
\end{tabular} 
\\
$^a$ Assuming a characteristic stellar mass of 0.5~M$_\odot$ as in \citet{Merin2008}.  \\
$^b$ Assuming the standard form of the IMF by \citet{Kroupa2005}.  \\
%$^c$ Assuming a characteristic stellar mass of 1.3~M$_\odot$ (see Sect.~\ref{LF}).  \\
\end{table*}

\begin{table*}
\small \caption{Total cloud mass (M$_{cloud}$) above the star formation threshold (A$_V \approx$8.6~mag) 
and fraction of M$_{cloud}$ below this threshold  for the Lupus clouds observed with Spitzer.  \label{mass_tab}}
\begin{tabular}{lll}\hline\hline
Region & M$_{cloud}$  above & Fraction of M$_{cloud}$ \\
              & threshold(M$_\odot$)      &     below threshold                  \\
\hline
Lupus~I       &     30          &   94\%    \\
Lupus~III     &     84          &   91\%     \\
Lupus~IV    &     47          &  75\%        \\
Lupus~V     &     0            & 100\% \\
Lupus~VI    &     2.3        &   99\%  \\
\\\hline
\end{tabular}
\end{table*}

\begin{table*}
\small \caption{Number and properties of the OB stars within 1~pc of each Lupus cloud observed by Spitzer. \label{OB_tab}}
\begin{tabular}{lllllll}\hline\hline
Region & N. of OB stars & RAJ2000  & DECJ2000  &  Spectral &  Luminosity     &  Projected distance   \\
              & within 1~pc                & (hh:mm:ss) &  (dd:mm:ss) &  Type      &    (L$_\odot$)  &      to the cloud            \\
\hline
Lupus~I       &  1 &  15:42:41.0    &  -34:42:36      & B5  &  163    & 36$\arcmin$ (0.65~pc) \\
Lupus~III     &  0 &  --			 &	--		 &  --    & --  &   --                                      \\   
Lupus~IV    &  0 &  --		 	 &	--		 &  --    & --  &    --                                     \\  
Lupus~V     &  1 &  16:24:31.8    & -37:33:58       & B8  &   119   &  44$\arcmin$ (0.96~pc) \\
Lupus~VI    &  1 & 16:26:20.5     &  -39:32:27      & B     &   2-3   &  46$\arcmin$ (1.00~pc) \\
\\\hline
\end{tabular}
\end{table*}

%%%%%%%%%%%%%%%%%%%%%%%%%%%

\section{Summary}
\label{summary}

We presented observations of the young stellar populations in the Lupus V and VI star-forming clouds
at 3.6, 4.5, 5.8, 8.0, 24.0 and 70.0 $\mu$m obtained with the Spitzer Space Telescope
IRAC and MIPS cameras and discussed them along with optical/near-infrared data available from
the literature. This study has been conducted within the frame of the Spitzer GB Survey.

The main results of this study are as follows:

\begin{itemize}

\item We found 43 YSO candidates in Lupus~V and 45 in Lupus~VI. None of them was 
classified as a PMS star from previous optical, near-IR and X-ray surveys;

\item Most of these YSOs candidates are surrounded by thin disks as deduced
from the high fraction of Class III objects (~79 \% in Lupus V,
~87\% in Lupus VI). This frequency of Class~III objects is significantly higher than measured
in other clouds of the Lupus complex and in other c2d clouds;

\item We found 2 potential transition objects, namely  ID~3 in Lupus V and ID~19 in Lupus VI. 
%Thus, the frequency of transitional objects in these clouds is about 2\%, 
%slightly smaller than measured in other YSO populations identified by Spitzer \citep[4-12\%;][]{Merin2010}.

\item The cloud density structure, the spatial distribution of the YSO candidates and the SFE in Lupus V and VI are similar 
to those of Lupus I and IV, but  clearly differ from those of Lupus~III.
Lupus V and VI present both a much lower SFE and a more spatially spread population of Class III YSOs with respect to Lupus~III. 
This suggests that the YSO population in Lupus V and VI might be a few Myrs older than the YSO population in Lupus~III.

\end{itemize}

We investigate possible scenarios explaining the high frequency of Class III YSO candidates in Lupus V and VI. 
We prove that contamination by field stars and disk photo-evaporation due to nearby OB stars can not be responsible for the observed high fraction of class III objects, 
while the higher characteristic stellar mass of YSO candidates in Lupus V and VI with respect to the other Lupus clouds may  be a contributing factor. 
Based on star formation rate and gas surface densities measurements, 
we conclude that star formation in Lupus V and VI ceased almost entirely a few Myrs ago, while this is not the case for other Lupus clouds. 
This is the most likely explanation for the observed high fraction of Class III objects. 
However, further observations allowing to put better constraints on the age and binary fraction 
of individual Lupus clouds would be necessary to definitively solve the puzzle.

\begin{acknowledgements}

We thank C. Reed for making available his all-sky catalog of OB stars,  A. Heiderman for providing unpublished calculations 
of the cloud mass in the Lupus complex and V. Roccatagliata and M. Gennaro for many useful discussions. 
We are also grateful to the anonymous referee for the useful revision of the paper, which helped us to improved its quality. 
Support for this work, part of the Spitzer Legacy Science Program, was
provided by NASA through contract number 1298236 issued by
the Jet Propulsion Laboratory, California Institute of Technology. 
Neal J. Evans  acknowledges support from NSF Grant AST-0607793.

\end{acknowledgements}

%%%%%%%%%%%%%%%%%%%%%%% Appendix

\appendix
\section{Appendix~A \label{appendixA}}

In this section we describe three alternative methods for YSO classification that we used to confirm 
the Lada classification based on the computation of the single $\alpha_{K-24}$ SED slope (Sect.~\ref{SEDs}).

\subsection{Stellar to disk luminosity ratio}

The ratio between the stellar and the disk luminosity gives information on the evolutionary status of the disk, i.e.
accreting, passive or debris disk \citep{Kenyon1987}.

\begin{figure*}
\includegraphics[angle=0,scale=0.9]{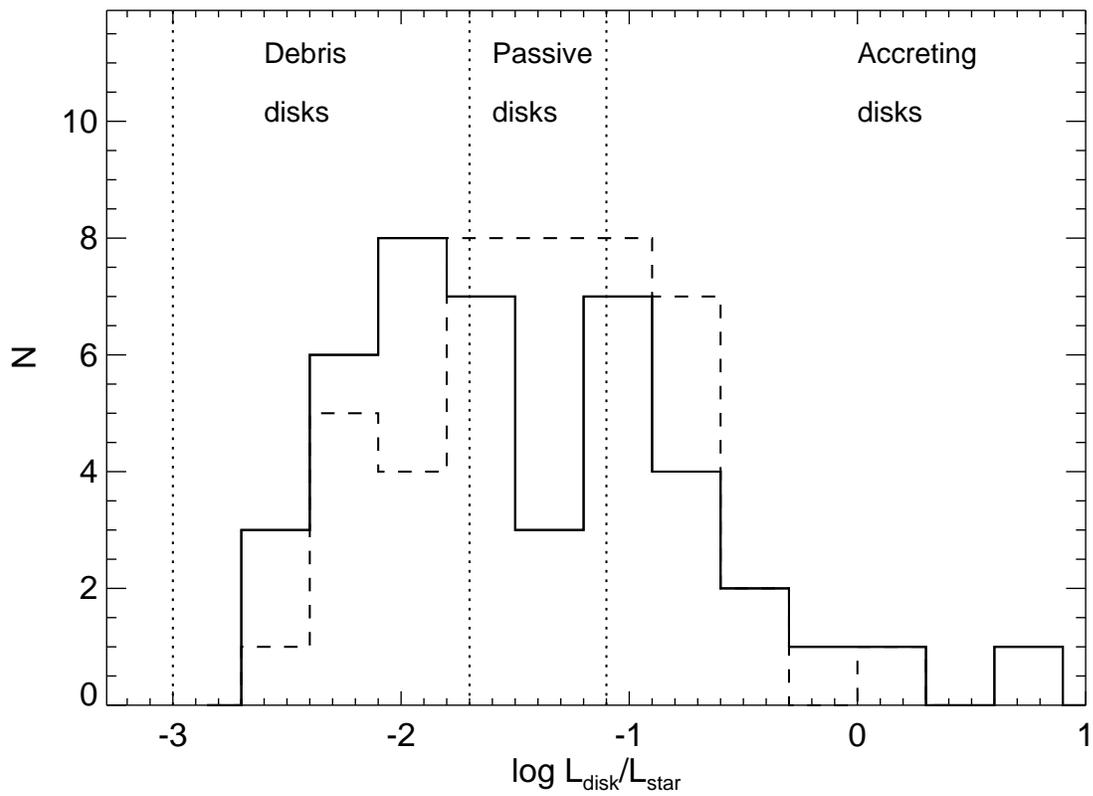}
\caption{Distribution of disk fractional luminosities for YSO candidates in Lupus V (solid line) and Lupus VI (dashed line). The vertical
dot-lines mark the typical boundaries of active, passive, and
debris disks as defined by \cite{Kenyon1987} \label{lumi}.}
\end{figure*}

We calculate the total luminosity (L) of the each YSO candidate by integrating under
all the available fluxes of the SED, as explained in Sect.~\ref{LF}, and
the stellar luminosity (L$_\star$) by integrating under the best fitting NextGen
photosphere model (Sect.~\ref{SEDs}); the difference between L and L$_\star$
gives the disk luminosity (L$_{disk}$).
Figure~\ref{lumi} shows that the disk fractional luminosities for the YSO candidates in
Lupus V and VI is such that they are mainly surrounded by
passive or debris disk, which confirms for most of them the Class III classification obtained from the $\alpha_{K-24}$-slope.

\subsection{Color-color and color-magnitude diagrams}

Infrared CC and CM diagrams are good diagnostic tools for investigating the evolutionary phase of circumstellar matter 
around YSOs \citep{Hartmann2005,Lada2006}.

\begin{figure*}
\includegraphics[angle=0,scale=0.5]{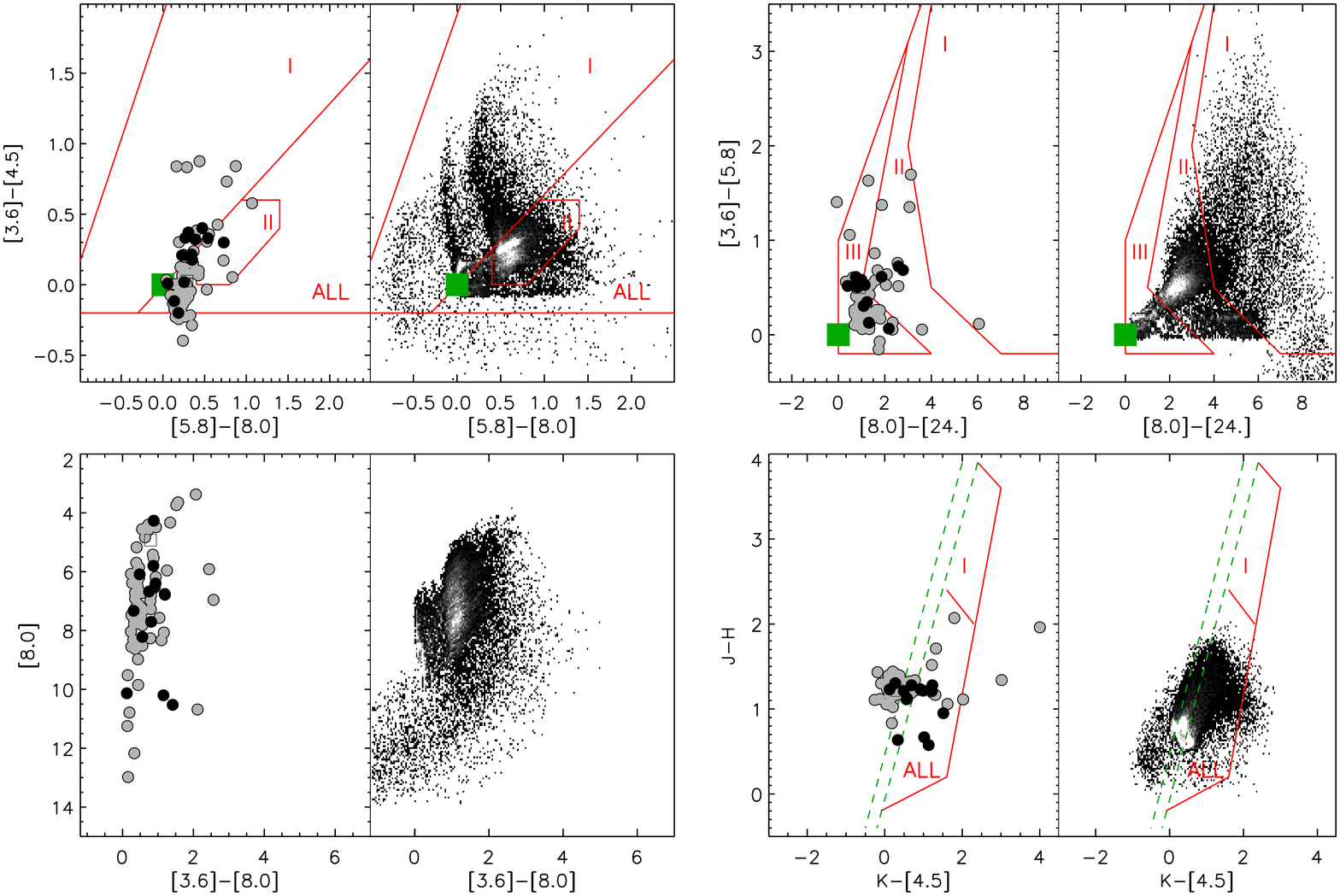}
\caption{In each panel, the left sub-panel shows the CC or CM diagram for the combined YSO population in Lupus V and VI
(black dots are the Class II YSO candidates, while gray dots ate the Class III YSO candidates), while the right sub-panel shows the colors
derived from the SED models by \cite{Robitaille2006}, in grayscale
intensity representing the density of points. The areas
corresponding to Stages I, II, and III as defined by
\cite{Robitaille2006} are only shown in some sub-panels. 
The label ``ALL'' marks the regions where models of all evolutionary
stages can be present. The big square and dashed-lines represent the location expected for normal
and reddened photospheres, respectively. \label{cccmd1}}
\end{figure*}

Figure~\ref{cccmd1} shows CC and CM diagrams in the Spitzer and 2MASS bands
for the YSO candidates in Lupus V and VI. The diagrams are compared with
those derived from the grid of YSO models by \cite{Robitaille2006},
adjusted in order to fit the distance of Lupus and matching the sensitivity
cutoff of the 2MASS \citep{Cutri2003} and Spitzer bands
\citep{Harvey2006,Jorgensen2006}.
This comparison shows that the majority of the YSO candidates in both Lupus V
and VI have colors consistent with Stage III objects in the \cite{Robitaille2006} scheme,
which is consistent with them being Class III objects in the Lada classification \citep[see also][]{Evans2009a}. 
This cross-check endorses both our classification and the reliability of the \cite{Robitaille2006} grid of disk models.

\subsection{SED morphology parameters \label{sed_par}}

An interesting diagnostic to study disk evolution in young low-mass
objects is the diagram of the SED excess slope, $\alpha_{excess}$,
versus the wavelength at which the IR excess begins, $\lambda_{turn-off}$
\citep[see][for the definition of both parameters]{Cieza2007,Harvey2007b,Harvey2008} .
Briefly, $\lambda_{turn-off}$ measures how far the circumstellar material extends inward to the
central star, while $\alpha_{excess}$ measures how optically thick the disk is.

\begin{figure*}
\includegraphics[angle=0,scale=0.9]{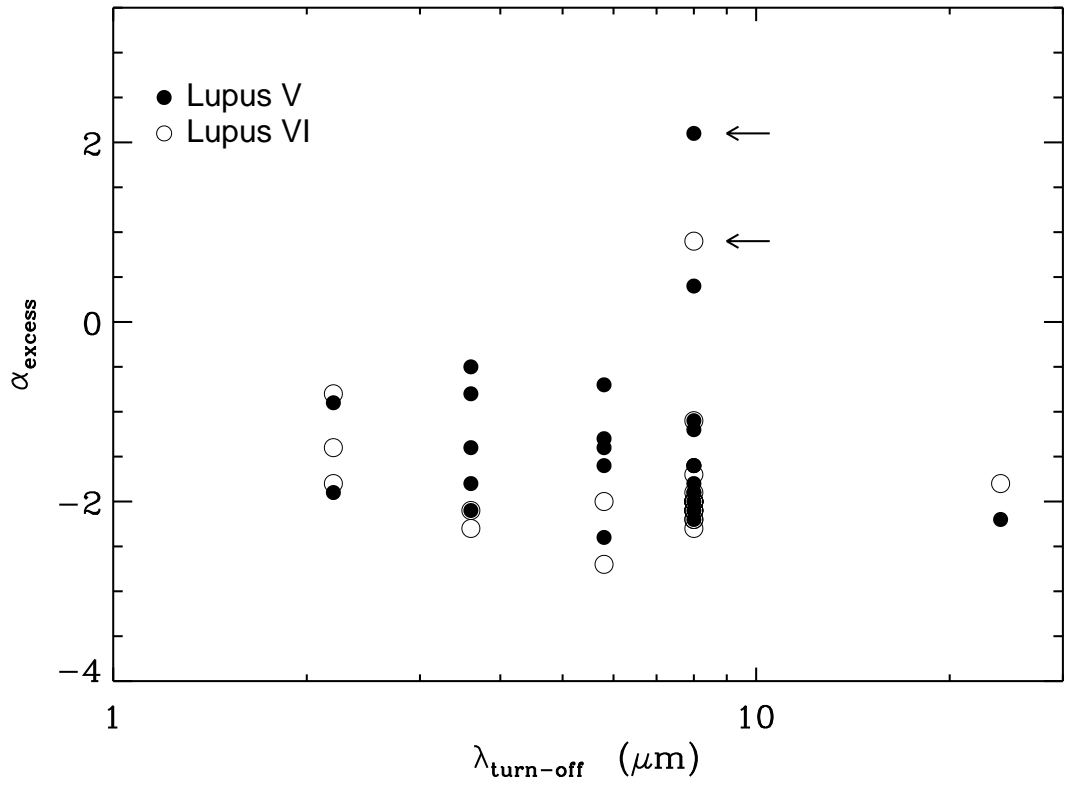}
\caption{SED excess slopes ($\alpha_{excess}$) as a function of the the wavelength at which the IR excess begins
($\lambda_{turn-off}$) for the Lupus V  and Lupus VI YSO candidates. 
The two arrows indicate the transitional object candidates. \label{alphaturnoff}}
\end{figure*}

Figure~\ref{alphaturnoff} shows such diagram for our sample of YSO candidates in Lupus V
and Lupus VI. Consistently with previous findings \citep[i.e.,][]{Cieza2007,Harvey2008}, 
the dispersion of $\alpha_{excess}$ starts increasing at $\lambda_{turn-off} \sim$4-5~$\mu$m,
which corresponds to a distance of about 0.2~AU from the central object
under the assumption of thermal emission from a thin disk with blackbody grains
\citep{Alcala2008}. In other words, disks in later evolutionary phases
(i.e. larger $\lambda_{turn-off}$) show a wider range of SED morphology (i.e. spread in $\alpha_{excess}$).
The the majority of the YSO candidates in both Lupus V and VI have $\lambda_{turn-off} \gtrsim$4~$\mu$m
indicative of evolved disks. Again, this is consistent with them being mainly Class III objects. 
The two transitional object candidates (ID~3 in Lupus V and ID~19 in Lupus VI) lie in the upper-left region of this diagram, 
as expected for disks with large inner holes \citep{Cieza2007}.

%%%%%%%%%%%%%%%%%%%%%%%%%%%%%%

\end{document}